\documentclass[aps,prl,twocolumn,superscriptaddress,groupedaddress]{revtex4-1}
\usepackage[compat=1.0.0]{tikz-feynman}
\pdfoutput=1

\usepackage{amssymb}
\usepackage{multirow}
\usepackage{bm}
\usepackage{amsmath}
\usepackage{graphicx}
\usepackage{epstopdf}
\usepackage{subfigure}
\usepackage{natbib}
\usepackage{epsfig}
\usepackage{amsfonts}
\usepackage{mathrsfs}
\usepackage[toc,page,title,titletoc,header]{appendix}
\usepackage[colorlinks,linkcolor=blue,citecolor=blue,anchorcolor=blue]{hyperref}

\usepackage{dsfont,amsthm,amsbsy}

\usepackage{fancyhdr}

\usetikzlibrary{quotes,angles}

\tikzset{-m-/.style={decoration={
  markings,
  mark=at position .65 with {\arrow{Stealth[width=2mm]}}},postaction={decorate}}}

\tikzset{->-/.style={decoration={
  markings,
  mark=at position .75 with {\arrow{Stealth[width=1.2mm]}}},postaction={decorate}}}

\newcommand{\smallplaquette}{\kern.08em
\begin{tikzpicture}[baseline={([yshift=-0.4ex]current bounding box.center)}]
    \draw coordinate (a) at (0,0);
    \draw coordinate (b) at (0.15,0);
    \draw coordinate (c) at (0.15,0.15);
    \draw coordinate (d) at (0,0.15);
    \draw (a) -- (b) -- (c) -- (d) -- (a) pic [draw=black]{} ;
\end{tikzpicture}%
\kern.08em%
}

\newcommand{\plaquette}{\kern.08em
\begin{tikzpicture}[baseline={([yshift=-.5ex]current bounding box.center)}]
    \draw coordinate (a) at (0,0);
    \draw coordinate (b) at (0.5,0);
    \draw coordinate (c) at (0.5,0.5);
    \draw coordinate (d) at (0,0.5);
    \draw (a) -- (b) -- (c) -- (d) -- (a) pic [draw=black, thin]{} ;
\end{tikzpicture}%
\kern.08em%
}

\newcommand{\plqzero}{\kern.08em
\begin{tikzpicture}[baseline={([yshift=-.5ex]current bounding box.center)}]
    \draw coordinate (a) at (0,0);
    \draw coordinate (b) at (0.5,0);
    \draw coordinate (c) at (0.5,0.5);
    \draw coordinate (d) at (0,0.5);
    \draw[->-, thin] (a) -- (b);
    \draw[->-, thin] (b) -- (c);
    \draw[->-, thin] (d) -- (c);
    \draw[->-, thin] (a) -- (d);
\end{tikzpicture}%
\kern.08em%
}

\newcommand{\plqone}{\kern.08em
\begin{tikzpicture}[baseline={([yshift=-.5ex]current bounding box.center)}]
    \draw coordinate (a) at (0,0);
    \draw coordinate (b) at (0.5,0);
    \draw coordinate (c) at (0.5,0.5);
    \draw coordinate (d) at (0,0.5);
    \draw[->-, thin] (b) -- (a);
    \draw[->-, thin] (b) -- (c);
    \draw[->-, thin] (d) -- (c);
    \draw[->-, thin] (d) -- (a);
\end{tikzpicture}%
\kern.08em%
}

\newcommand{\plqtwo}{\kern.08em
\begin{tikzpicture}[baseline={([yshift=-.5ex]current bounding box.center)}]
    \draw coordinate (a) at (0,0);
    \draw coordinate (b) at (0.5,0);
    \draw coordinate (c) at (0.5,0.5);
    \draw coordinate (d) at (0,0.5);
    \draw[->-, thin] (b) -- (a);
    \draw[->-, thin] (b) -- (c);
    \draw[->-, thin] (d) -- (c);
    \draw[->-, thin] (a) -- (d);
\end{tikzpicture}%
\kern.08em%
}

\newcommand{\plqthree}{\kern.08em
\begin{tikzpicture}[baseline={([yshift=-.5ex]current bounding box.center)}]
    \draw coordinate (a) at (0,0);
    \draw coordinate (b) at (0.5,0);
    \draw coordinate (c) at (0.5,0.5);
    \draw coordinate (d) at (0,0.5);
    \draw[->-, thin] (a) -- (b);
    \draw[->-, thin] (b) -- (c);
    \draw[->-, thin] (c) -- (d);
    \draw[->-, thin] (d) -- (a);
\end{tikzpicture}%
\kern.08em%
}

\newcommand{\plqthreereverse}{\kern.08em
\begin{tikzpicture}[baseline={([yshift=-.5ex]current bounding box.center)}]
    \draw coordinate (a) at (0,0);
    \draw coordinate (b) at (0.5,0);
    \draw coordinate (c) at (0.5,0.5);
    \draw coordinate (d) at (0,0.5);
    \draw[->-, thin] (b) -- (a);
    \draw[->-, thin] (c) -- (b);
    \draw[->-, thin] (d) -- (c);
    \draw[->-, thin] (a) -- (d);
\end{tikzpicture}%
\kern.08em%
}

\newcommand{\vertex}{\kern.08em
\begin{tikzpicture}[baseline={([yshift=-.5ex]current bounding box.center)}]
    \draw coordinate (o) at (0,0);
    \draw coordinate (x) at (0.35,0);
    \draw coordinate (mx) at (-0.35,0);
    \draw coordinate (z) at (0,0.35);
    \draw coordinate (mz) at (0,-0.35);
    \draw coordinate (y) at (0.28,0.21);
    \draw coordinate (my) at (-0.28,-0.21);
    \draw[ultra thin] (o) -- (x);
    \draw[ultra thin] (o) -- (mx);
    \draw[ultra thin] (o) -- (y);
    \draw[ultra thin] (o) -- (my);
    \draw[ultra thin] (o) -- (z);
    \draw[ultra thin] (o) -- (mz);
\end{tikzpicture}%
\kern.08em%
}

\newcommand{\smallvertex}{\kern.08em
\begin{tikzpicture}[baseline={([yshift=-.5ex]current bounding box.center)}]
    \draw coordinate (o) at (0,0);
    \draw coordinate (x) at (0.15,0);
    \draw coordinate (mx) at (-0.15,0);
    \draw coordinate (z) at (0,0.15);
    \draw coordinate (mz) at (0,-0.15);
    \draw coordinate (y) at (0.12,0.09);
    \draw coordinate (my) at (-0.12,-0.09);
    \draw[thin] (o) -- (x);
    \draw[thin] (o) -- (mx);
    \draw[thin] (o) -- (y);
    \draw[thin] (o) -- (my);
    \draw[thin] (o) -- (z);
    \draw[thin] (o) -- (mz);
\end{tikzpicture}%
\kern.08em%
}

\newcommand{\vertexcoplanar}{\kern.08em
\begin{tikzpicture}[baseline={([yshift=-.5ex]current bounding box.center)}]
    \draw coordinate (o) at (0,0);
    \draw coordinate (x) at (0.35,0);
    \draw coordinate (mx) at (-0.35,0);
    \draw coordinate (z) at (0,0.35);
    \draw coordinate (mz) at (0,-0.35);
    \draw coordinate (y) at (0.28,0.21);
    \draw coordinate (my) at (-0.28,-0.21);
    \draw[->-, ultra thin] (o) -- (x);
    \draw[->-, ultra thin] (o) -- (mx);
    \draw[->-, ultra thin] (o) -- (y);
    \draw[->-, ultra thin] (my) -- (o);
    \draw[->-, ultra thin] (z) -- (o);
    \draw[->-, ultra thin] (mz) -- (o);
\end{tikzpicture}%
\kern.08em%
}

\newcommand{\vertexnoncoplanar}{\kern.08em
\begin{tikzpicture}[baseline={([yshift=-.5ex]current bounding box.center)}]
    \draw coordinate (o) at (0,0);
    \draw coordinate (x) at (0.35,0);
    \draw coordinate (mx) at (-0.35,0);
    \draw coordinate (z) at (0,0.35);
    \draw coordinate (mz) at (0,-0.35);
    \draw coordinate (y) at (0.28,0.21);
    \draw coordinate (my) at (-0.28,-0.21);
    \draw[->-, ultra thin] (o) -- (x);
    \draw[->-, ultra thin] (mx) -- (o);
    \draw[->-, ultra thin] (o) -- (y);
    \draw[->-, ultra thin] (my) -- (o);
    \draw[->-, ultra thin] (o) -- (z);
    \draw[->-, ultra thin] (mz) -- (o);
\end{tikzpicture}%
\kern.08em%
}

\def\Ueph{U_\text{e-ph}}

\begin{document}

\title{Emergent Gauge Fields in Band Insulators }

\author{Zhaoyu~Han}
\affiliation{Department of Physics, Stanford University, Stanford, California 94305, USA} 

\author{Steven~A.~Kivelson} 
\email{kivelson@stanford.edu}
\affiliation{Department of Physics, Stanford University, Stanford, California 94305, USA}

\date{\today}

\begin{abstract}
    By explicit microscopic construction involving a mapping to a quantum vertex model subject to the `ice rule,' we show that an electronically `trivial' band insulator with suitable vibrational (phonon) degrees of freedom can host a ``resonating valence-bond'' state - a quantum phase with emergent gauge fields. This novel type of band insulator is identifiable by the existence of emergent gapless `photon' modes and deconfined excitations, the latter of which carry non-quantized mobile charges. We suggest that such phases may exist in the quantum regimes of various nearly ferroelectric materials.
\end{abstract}

\maketitle

{\bf Significance Statement: The ice rule constrains the bond configuration in water ice and spin ice materials, which has profound implications for their properties, such as allowing phases with emergent gauge fields. In this work, we construct a simple electron-phonon model with an exact ice rule when the coupling strength is strong enough and prove the possibility of a resonating valence bond phase in its phase diagram. With this concrete example, we hope to expand the range of physical systems suitable for fractionalization phenomena to include band insulators with relatively strong electron-phonon couplings.}

The term ``resonating valence-bond (RVB) state,'' initially introduced by Pauling in 1949~\cite{doi:10.1098/rspa.1949.0032}, refers to phases of matter in which the quantum degrees of freedom fluctuate in a coherent way. The idea was revisited by Anderson~\cite{ANDERSON1973153,doi:10.1126/science.235.4793.1196} in the context of frustrated quantum antiferromagnets, inspiring extensive theoretical investigations of ``spin liquids''~\cite{savary2016quantum,doi:10.1126/science.aay0668,RevModPhys.89.025003,doi:10.1126/science.1163196} in ``Mott insulators,'' i.e. interaction dominated insulators with an odd number of electrons per unit cell,  topological degeneracies or gapless neutral excitations (due to the Lieb-Schultz-Mattis theorem~\cite{LIEB1961407,PhysRevLett.84.1535,PhysRevB.69.104431}), and fractionalized excitations~\cite{PhysRevB.35.8865,PhysRevLett.59.2095}. From a modern perspective, gapped spin liquids belong to the class of `topologically ordered' quantum phases~\cite{doi:10.1142/S0217979290000139}, whose universal properties have been mostly classified and characterized~\cite{RevModPhys.89.041004,sachdev2023quantum}.

Despite the successes of the abstract theory, it remains somewhat disappointing that while there are an increasing number of promising ``spin liquid candidate materials,''~\cite{annurev:/content/journals/10.1146/annurev-matsci-080819-011453,chamorro2020chemistry,TREBST20221} there is no single solid material that has been indisputably established as having an RVB groundstate. Expanding the search for RVB to a broader range of systems is thus desirable, given that the essential feature of an RVB phase is the emergence of gauge fields at low energy~\cite{PhysRevB.44.2664,PhysRevB.44.2664,PhysRevLett.66.1773,PhysRevB.62.7850, PhysRevLett.91.167004, PhysRevB.68.184512,KITAEV20062,PhysRevB.68.184512, PhysRevB.69.224415,PhysRevB.69.224416,  RevModPhys.78.17,PhysRevB.65.024504, PhysRevB.67.245316,moessner2010quantum, fradkin2013field,sachdev2016emergent,sachdev2018topological} which does not necessarily have any relation to spin physics and/or Mott insulators. Indeed, proposals for finding RVB have been recently pursued in Rydberg atom arrays~\cite{doi:10.1126/science.abi8794,PhysRevX.11.031005,PhysRevX.12.041029,PhysRevLett.130.043601,cheng2024emergent,shah2023quantum,cheng2024emergent} and other quantum engineering platforms~\cite{PhysRevLett.89.277004,ioffe2002topologically,pinto2020gauge,bǎzǎvan2024synthetic}.

\begin{figure}[t]
    \centering
    \includegraphics[width = 0.8 \linewidth]{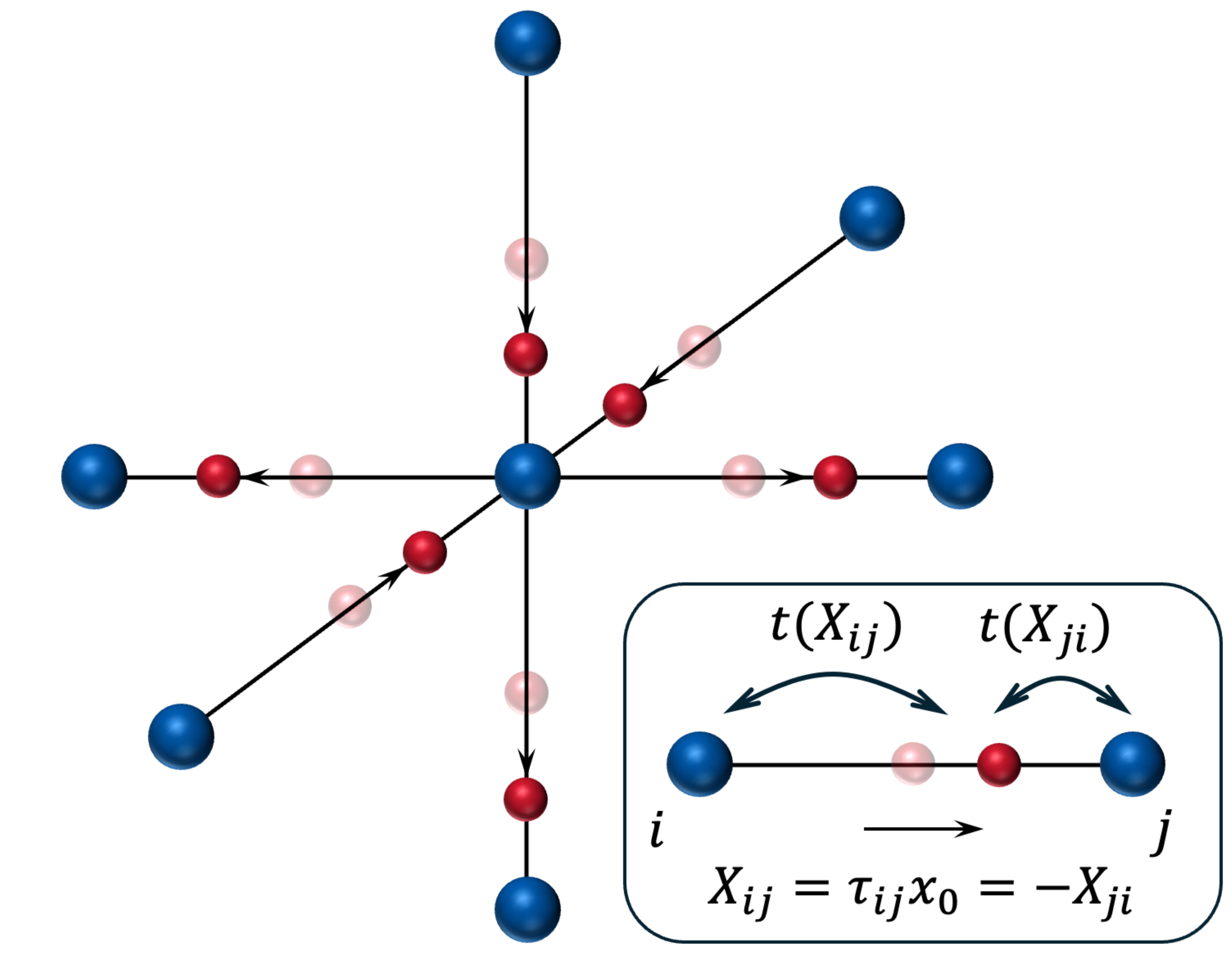}
    \caption{An illustration of the lattice structure around a lattice vertex and the ice-rule constraint: there are the same numbers of (red) bond atoms displaced toward as away from each (blue) vertex atom. Inset:  the parametrization of a bond atom's displacement and its effect on the electron hopping amplitude.
}
    \label{fig: illustration}
\end{figure}

\begin{figure}[t]
    \centering
    \includegraphics[width = \linewidth]{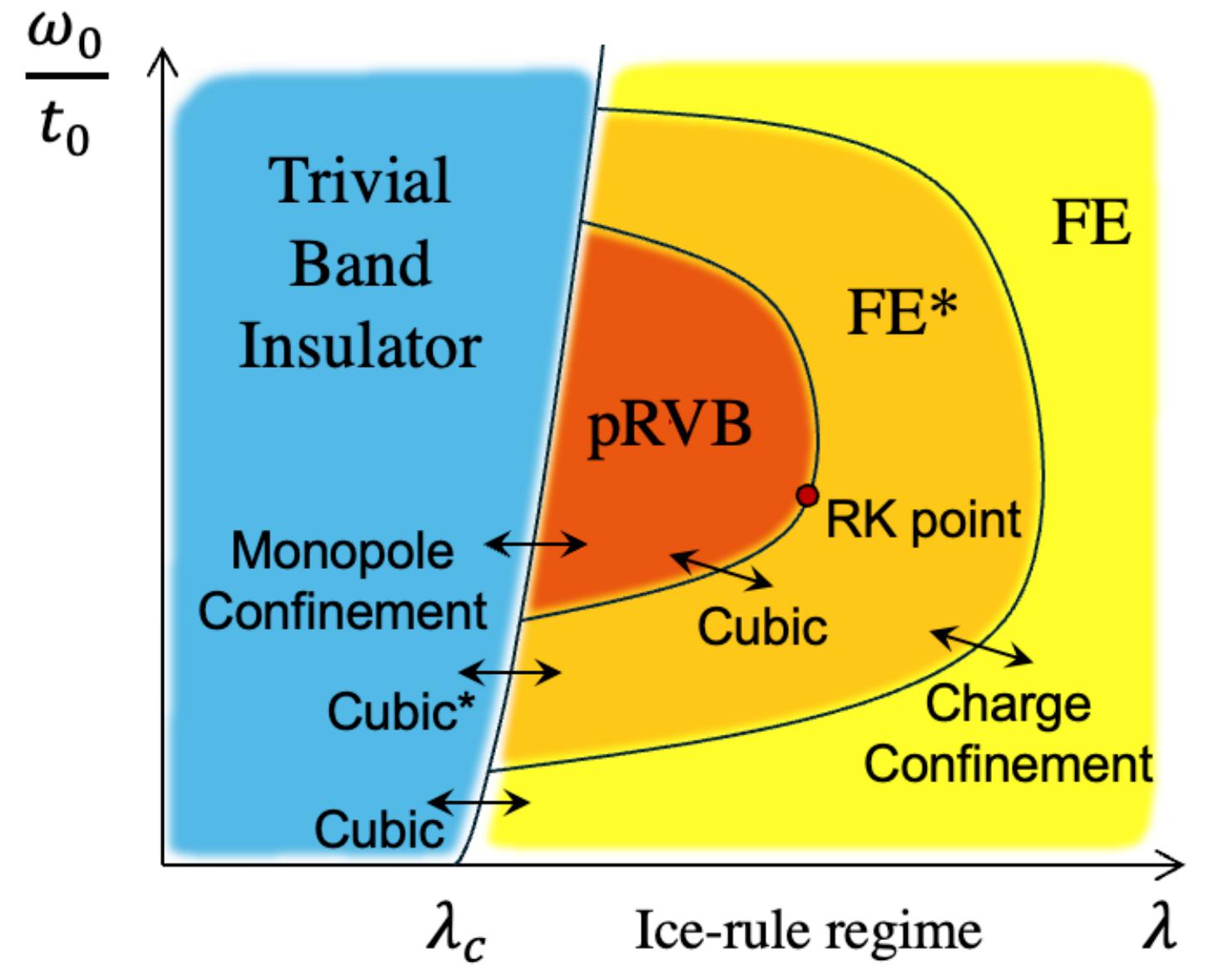}
    \caption{A schematic representation of one of the possible phase diagrams of the model in Eq.~\ref{eq: microscopic model} on a 3D cubic lattice. $\lambda =\Ueph /t_0$ is a dimensionless measure of the strength of the EPC, i.e., the ratio between the EPC-induced interaction and the bare hopping amplitude. The nature of the various phases and phase transitions is discussed in the text and Sec.~IV of SM.}
    \label{fig: phase diagram}
\end{figure}

In this work, we articulate a different route to a quantum fluid phase with an emergent gauge field, which builds on phonon degrees of freedom on lattice bonds, roughly realizing the original intuitive picture of RVB. We note that certain aspects of this perspective have been discussed in studies of water ice~\cite{PhysRevB.93.125143,PhysRevB.99.174111,henley2010coulomb,PhysRevB.100.014206,lin2011correlated,10.1063/1.1681056,PhysRevB.93.125143,PhysRevB.74.024302,de2020crystal} and spin ice~\cite{doi:10.1146/annurev-conmatphys-031218-013401,Gingras_2014,castelnovo2012spin,PhysRevLett.112.167203,PhysRevB.68.115413,PhysRevB.108.L220402,castelnovo2008magnetic}. To distinguish such states from the spin liquids discussed in much literature, we refer to them as phononic resonating-valence-bond (pRVB) states. Previously, we have shown with two explicit examples~\cite{PhysRevLett.130.186404,cai2024quantum} that certain models with strong electron-phonon couplings (EPC) and one electron per unit cell can exhibit RVB phases (i.e., Mott physics). Here, we consider the effect of another type of strong EPC in a band-insulator, where there is an even number of electrons per unit cell and always a gap in the electronic spectrum, even in the absence of EPC. We introduce an explicit minimal microscopic model (Eq.~\ref{eq: microscopic model}) that captures the vibrational effects of atoms on lattice bonds, and show that when the strength of the EPC exceeds a certain threshold, the low-energy physics can be well described by a quantum vertex model whose Hilbert space is constrained by a (generalized) ice rule: around each lattice vertex, there are the same number of surrounding bond atoms that move closer to and away from it. (See Fig.~\ref{fig: illustration} for illustration.) This constraint is a highly non-perturbative consequence of the electrons' effects on the phonons, as we establish using both exact arguments and numerical extensions. Then, by analyzing an exactly solvable point of the effective quantum vertex model and an effective field theory that is valid in the neighborhood of this point, we establish the possible existence of several phases in the phase diagram in three dimensions (see Fig.~\ref{fig: phase diagram}): a pRVB phase featuring an emergent $U(1)$ gauge field, a ferroelectric (FE) phase, and a phase with both FE order and an emergent gauge field that we refer to as FE$^*$.

While our simple model does not directly apply to any specific material we know of, it is a plausible caricature of various nearly ferroelectric materials including, possibly, water ice at suitable pressures. We thus suggest that such materials should be reconsidered and reinvestigated in the context of ``pRVB candidate materials.''  To experimentally distinguish pRVB states from trivial band insulators, we analyze their identifying features, including gapless transverse optical (TO) phonons (emergent photons) and, at energy scales well below the electronic gap, deconfined excitations with a non-quantized charge. The latter give rise to an anomalously high conductivity at temperatures well below the electronic band gap.

\section{The microscopic model} 

We now introduce a model that captures the effects of the motion of bond-centered atoms parallel to the bonds' direction. This model can be defined in any dimension on any lattice in which all the bonds are equivalent under the space group symmetry, and each bond perpendicularly intersects a (glide) reflection plane or a 180\textdegree-rotation axis at the bond center. This includes a huge family of lattices, e.g., simple, body-centered, or face-centered cubic, and diamond lattices in three dimensions. Given such a lattice, we consider its Lieb-type~\cite{lieb2004two} generalization by putting an extra atomic orbital on each bond center.  The minimal model then reads:
\begin{align}\label{eq: microscopic model}
    \hat{H} =& \sum_{i}\left[\Delta\ \hat{n}_i - \sum_{j \in \text{nn\ } i, \sigma} t(\hat{X}_{ij})\left(\hat{c}^{\dagger}_{i\sigma} \hat{f}_{\langle ij \rangle  \sigma}+\text{h.c.}\right) \right] \nonumber\\
    & \ \ \ + \sum_{\langle ij \rangle} \left(\frac{K\hat{X}_{ij}^2}{2}+\frac{\hat{P}_{ij}^2}{2M}\right) + \dots
\end{align}
where $\hat{c}_{i\sigma}$ and $\hat{f}_{\langle ij \rangle\sigma}$ are the electron annihilation operators on vertex $i$ and nearest-neighbor bond $\langle ij\rangle$, respectively, $\hat{n}_i$ and $\hat{n}_{\langle ij\rangle}$ are the corresponding electron densities, $\Delta$ is the charge transfer gap between vertex and bond orbitals, and $\text{nn\ }i$ stands for the nearest neighboring vertices around vertex-$i$. $\hat{X}_{ij}$ is the displacement of the bond atom relative to the center of bond $\langle ij\rangle$ in the direction of $i$-to-$j$ - i.e. $X_{ij}=-X_{ji}$ with the origin defined such that $X_{ij}=0$ when the bond atom is at the bond center, and $\hat{P}_{ij}$ is the conjugate momentum operator. They describe the out-of-phase motion of the bond atoms relative to the site atoms and assumed to be optical phonons with flat bare dispersion $\omega_0\equiv \sqrt{K/M}$, where $K$ is the bare stiffness, and $M$ is the bond atom mass. The symbol $\dots$ represents the additional physically plausible terms that we will first neglect and later treat perturbatively.

\section{The  emergence of the ice rule}

In the classical limit of the problem, where $M\rightarrow \infty$, the energy of the system is a function of the phonon configuration, $E^\text{cl}(\{X_{ij}\})$, which consists of the phonon elastic energy plus the sum of the electron energies in all occupied states - i.e. all states with energy below the Fermi level.  This quantity is a complicated, non-local function of $\{X_{ij}\}$ for a general phonon configuration. However, as shown in Sec.~I~A of supplemental materials (SM), when the lattice coordination number is even, we prove a remarkable feature of the energy function:

{\bf Theorem I:} {\it For any electron density, $E^\text{cl}$ is the same for all phonon configurations in which the magnitudes of the atomic displacements are equal on all bonds,  i.e., $|X_{ij}|=x_0$, and  which satisfy an ``ice rule'' constraint at every vertex, i.e. $\sum_{j\in \text{nn }i} X_{ij}=0$.}

We call these ``ice-rule (satisfying) 
configurations'' as this is a precisely realized version of the kind of (approximate) degeneracy of proton configurations seen in water ice. Any such $\{X_{ij}\}$ can be parameterized by the direction of polarization on each bond, i.e., $X_{ij} = \tau_{ij} x_0 $ with $\tau_{ij}=-\tau_{ji} = \pm 1$ being oriented pseudospin variables, which in turn satisfy the constraints $\sum_{j\in \text{nn }i} \tau_{ij}=0$. The above theorem implies that, for ice-rule configurations, $E^\text{cl}$ generally depends on the amplitude $x_0$ but is independent of $\{\tau_{ij}\}$:
\begin{align}
    E^{\text{cl}}(\{X_{ij}\}) = E^\text{ice} (x_0).
\end{align}

{\bf Addendum:} {\it Not only the classical ground-state energy but also the entire electronic spectra are exactly the same for all ice-rule configurations with the same $x_0$. }

We emphasize the above statement is exact, regardless of the form of the function $t(x)$. This derives from the lattice symmetries we assumed and the special structure of the hopping matrix (i.e., there is no direct hopping between $f$ orbitals). Therefore, it not only holds for the specific Hamiltonian in Eq.~\ref{eq: microscopic model} but pertains in the presence of various kinds of additional terms, such as direct hopping among $c$-orbitals, e.g. $\sum_{\langle ij\rangle}t'(\hat{X}_{ij}) (\hat{c}^\dagger_{i}\hat{c}_j +\text{h.c.})$, phonon coupling to the local electron densities, e.g. $\sum_{\langle ij\rangle}a(\hat{X}_{ij}) (\hat{n}_{i}-\hat{n}_j) + b(\hat{X}_{ij}) (\hat{n}_{i} + \hat{n}_j) + c(\hat{X}_{ij}) \hat{n}_{\langle ij\rangle} $, etc.

We have not proven that ice-rule configurations are the global minima of $E^\text{cl}$, nor have we obtained an analytic expression for the optimal value of $x_0$ for them (although this is easy to determine numerically). However, we have proven a partial result:

{\bf Theorem II:} {\it For any electron density, ice-rule configurations are stationary points of the classical energy manifold with a fixed root-mean-square displacement, $\sqrt{\overline{X_{ij}^2}}$.}

To confirm whether such configurations are the classical ground states at electron densities corresponding to band insulators, we have numerically investigated the model on large clusters of various one-, two- and three-dimensional regular lattices with two electrons per vertex, adopting the simple form $t(x) = t_0 - g x$, which is generic for small displacements (See Sec.~I~C of SM for results of the cubic lattice). We find that the ice-rule configurations (including the trivial case with $x_0=0$) are indeed always the lowest-energy ones at all sets of parameters we investigated; moreover, at generic values of the charge transfer gap $\Delta$, when the EPC-induced interaction strength $\Ueph \equiv g^2/K$ surpasses a certain threshold determined by $t_0$ and $\Delta$, the system undergoes a continuous transition from a trivial band insulator (in which $E^{cl}$ is minimized by $x_0=0$) to an ice-rule regime with $x_0\neq 0$. Specifically, defining a dimensionless factor $\lambda \equiv \Ueph/t_0$, we find that the optimal $x_0$ scales as $\sim \sqrt{\lambda -\lambda_c}$ as $\lambda$ approaches the critical value, $\lambda_c$, from the ice-rule regime. Meanwhile, the electronic spectrum is well-gapped on both sides of the transition with a gap $\sim |\Delta|$; thus, the system is always a `band insulator' if only the electron sector of the problem is concerned. 

\section{Effective Hamiltonian} 

Both the effects of non-infinite ion mass $M$ and other physically allowed couplings represented by $\dots$ in Eq.~\ref{eq: microscopic model} generically lift the ice-rule degeneracy. So long as the additional interactions are weak (in appropriate units), their effects can be computed perturbatively. For example, dynamically induced degeneracy-lifting terms (which may cause ``order by disorder'') can be computed in powers of $\omega_0$, starting with a leading term that encodes the configuration dependence of the zero-point energy of the phonons in the harmonic approximation. Moreover, non-zero $\omega_0$ also permits tunneling processes between distinct states in the degenerate ground-state manifold of $E^{cl}$; the amplitudes of such processes can be computed using semi-classical (instanton) methods. All these effects can be modeled by an effective Hamiltonian, $\hat H^\text{eff}$, that acts in the subspace of ice-rule configurations with basis states indexed by the direction of the dipole on each bond, $\{\tau_{ij}\}$. For the sake of concreteness, from here on, we focus on the case of a cubic lattice. 

Since the leading order terms in $\hat H^\text{eff}$ are expected to be short-ranged, we focus our attention on the shortest-range terms consistent with the constraints of lattice symmetry and the ice rule.  It is possible to express these as a sum of vertex terms - i.e., terms that operate on the bond variables radiating from a given vertex - and plaquette terms - i.e., those that depend on the bond variables surrounding an elementary plaquette. Fortunately, as shown in Sec.~II~B of SM, the ice-rule constraint makes the vertex terms redundant; the shortest-range model can thus be expressed entirely in terms of a sum over plaquette operators. The same analysis can be straightforwardly extended to other lattices and to include further neighbor interactions.

To most conveniently express the model, we pictorially represent the dipole variable $\tau_{ij}$ on each bond by an arrow and each square plaquette of the lattice by symbol $\plaquette$. A given plaquette can have any of $2^4$ possible edge configurations, but these can be grouped into four equivalent types under the lattice symmetries, whose representatives are, respectively, type-$0$: \plqzero, type-$1$:  \plqone, type-$2$: \plqtwo, and type-$3$: \plqthree. 
We denote the projector onto the subspace corresponding to type-$m$ configurations on a plaquette as $\hat{P}_m^{\smallplaquette}$. In terms of these, the effective model is (the projection onto the ice-rule space is left implicit)
\begin{align} \label{eq: effective model}
\hat{H}^\text{eff} =  \sum_{\smallplaquette} \left[\sum_{m=1,2,3} V_m  \hat{P}_m^{\smallplaquette} - J \left(\left|\plqthree \right\rangle \left\langle \plqthreereverse \right| + \text{h.c.}\right)\right]
\nonumber
\end{align}
where $V_m$ terms are the interactions that are diagonal in the $\tau$ basis ($V_0$ is set to $0$ without loss of generality), and $J$ is the amplitude of the minimal quantum tunneling process within the ice-rule subspace activated by finite $\omega_0$. 
This is a variant of quantum vertex models discussed in Refs.~\cite{ARDONNE2004493,PhysRevB.69.064404,PhysRevB.106.195127,balasubramanian2022exact}.

Because it derives from a tunneling process, $J\sim \mathrm{e}^{-S}$ where the tunneling action $S$ is proportion to $\sqrt{M}$, the tunneling distance $\sim x_0$, and the energy of the barrier to tunneling. 
Thus, whenever $\lambda$ is large enough to produce substantial dipolar displacements as well as tunneling barriers, $J$ is negligible compared to the interaction $V_m$'s.  In this limit, $\hat H^\text{eff}$ is effectively classical ($J\approx 0$), with ordered ground states that can be determined exactly.  Depending on the relative values of $\{V_m\}$, various phases are possible, including several distinct ferro-, antiferro- and ferri-electric orders, as we present in Sec.~II~C of SM. 

However, as $\lambda$ approaches $\lambda_c$, the action of the minimal quantum tunneling event vanishes in proportion to $x_0^3 \sim (\lambda-\lambda_c)^{3/2}$ (Sec.~I~D of the SM). This derives from a combination of an increasingly small tunneling distance ($\sim x_0$) and an increasingly small tunnel barrier $\sim t_0 x_0^4$. Thus, over a relatively broad range of circumstances not too far from criticality, where the magnitude of the dipolar distortions satisfies  $(\lambda-\lambda_c)^{3/2} \sim
(\omega_0/t_0)$, the model enters a highly quantum regime in which $J$ can be substantial and comparable to $V_m$.

\section{Effective field theory}

Fortunately, like its cousin - the quantum dimer model~\cite{PhysRevLett.61.2376} - this model has a so-called Rokhsar-Kivelson (RK) point at which the ground states are exactly solvable. This corresponds to  $V_3=J$, $V_1=V_2=0$, where each groundstate is an equal superposition of all ice-rule configurations within a topological sector specified by the total `fluxes' of $\tau_{ij}$ (viewing it as a vector field) through the non-contractible closed surfaces of the system. Also similar to the case in the quantum dimer model~\cite{PhysRevB.68.184512,PhysRevB.69.224415,moessner2010quantum,fradkin2013field} in three dimensions, the long-wavelength physics near the RK point is described by a $U(1)$ gauge theory with Hamiltonian density (Sec.~III~A of SM):
\begin{align}
    \hat{\mathcal{H}} = \frac{1}{2\tilde{\epsilon}}|\hat{\bm{d}}|^2 + \frac{1}{2\tilde{\mu}}\left(\kappa^2|\nabla\times\hat{\bm{d}}|^2 +|\nabla \times \hat{\bm{a}}|^2 \right)
\end{align}
where $\kappa$, $\Tilde{\epsilon}$, $\tilde{\mu}$ are constants ($\kappa\approx 3.6$ for the cubic lattice), $\bm{d}$ is the coarse-grained field version of $\tau_{ij}/2$ subject to the ice-rule constraint $\nabla \cdot \bm{d} =0$, and $\hat{\bm{a}}$ is the conjugate field operator of $\hat{\bm d}$ that satisfies the canonical commutation relations $[\hat{\bm{d}}_\alpha(\bm{r}),\hat{\bm{a}}_\beta(\bm{r}')] = \mathrm{i} \delta_{\alpha \beta} \delta(\bm{r}-\bm{r}')$. By analogy to the electromagnetic (EM) field, $\bm{d}$ can be viewed as an emergent electric displacement field, and $\bm{b}\equiv \nabla\times \bm{a}$ is thus the emergent magnetic flux density. (The emergent electric and magnetic fields $\bm{e}\equiv \bm{d}/\Tilde{\epsilon}$, $\bm{h}\equiv \bm{b}/\Tilde{\mu}$ are related to these fields by the effective `permittivity,' $\tilde{\epsilon}$, and `permeability,' $\tilde{\mu}$.) At the RK point, the theory is anomalous in the sense that $1/\tilde{\epsilon}=0$. The corresponding field theory describes an unstable fixed point, obtained when the amplitude ($1/\tilde{\epsilon}$) of the only relevant operator ($\hat{\bm{d}}^2$) vanishes~\cite{PhysRevB.69.224415}.

Tuning away from this point, when $1/\tilde{\epsilon}>0$, the second term ($|\nabla\times \hat{\bm{d}}|^2$) can be dropped in the infrared limit, and the theory becomes exactly the same as that describing the usual EM gauge field. In this way, we have theoretically established the existence of the pRVB phase with deconfined $U(1)$ gauge fields in the vicinity of the RK point of the model in Eq.~\ref{eq: effective model}. Indeed, numerical studies on a similar vertex model on a diamond lattice have verified such a phase near its RK point~\cite{PhysRevLett.108.067204,PhysRevLett.127.117205,PhysRevB.84.115129}. We note that the $U(1)$ gauge fields are {\it emergent}, in the sense that the microscopic theory does not have exact gauge invariance; nonetheless, their deconfined phase can still be stabilized~\cite{FOERSTER1980135}. 

On the other hand, when $1/\tilde{\epsilon} <0$, higher-order terms need to be included, following which we find a variety of possible broken symmetry phases with finite expectation values of $\langle \bm{d}\rangle$ (See Sec.~III~B of SM for details). These higher terms are dangerously irrelevant perturbations of the unstable fixed point at $1/\tilde{\epsilon} =0$, which is thus a multicritical point. For instance, in the presence of positive quartic terms ($|\bm{d}|^4$ and $\sum_{\alpha = x,y,z} \bm{d}_\alpha^4$), the polarization does not jump discontinuously to large values but rather rises continuously with increasing $|1/\tilde{\epsilon}|$. The corresponding phase is a novel FE$^*$ phase in which partial spontaneous polarization coexists with gapless emergent photons and deconfined excitations as in the $1/\tilde{\epsilon}>0$ phase. We sketch a possible phase diagram for this scenario in Fig.~\ref{fig: phase diagram}.

Like the native EM $U(1)$ gauge field, the emergent gauge field has gapless `photon' excitations with `speed of light' $\tilde{c}=1/\sqrt{\tilde{\epsilon}\tilde{\mu}}$, and couple to `charged' excitations. Microscopically, the gapless photons are TO phonon modes. The elementary charged excitation is a local violation of the ice rule around a vertex that has a net $\pm 1$ units of electric flux, i.e. $\oint d\bm{S} \cdot \bm{d}=\pm 1$ for any surface enclosing it. Moreover, since this is an emergent gauge field on a lattice, magnetic monopole excitations are also possible, and they are quantized due to Dirac's quantization. Both types of excitations are deconfined in the sense that two charges or monopoles interact at large distances via a Coulomb interaction. The transitions out of this deconfined Coulomb phase for the gauge theory can be understood as condensation of either the charges or the monopoles, which leads to the confinement of the other~\cite{PhysRevD.19.3682, fradkin2013field}. By tracking the fate of the excitations at the phase boundaries, we are able to deduce the nature of the transitions in Fig.~\ref{fig: phase diagram} with the caveat that they may be weakly first order due to fluctuation effects~\cite{PhysRevLett.32.292} (see Sec.~IV of SM for details). In particular, the transition between the pRVB and trivial band insulator phases may be understood as condensation of the charge excitations, whose creation energy vanishes as $(\lambda-\lambda_c)^2$ near the transition.

\section{Phenomenology of the pRVB phase}

If one is lucky enough to find a material with a pRVB or FE* ground state, this can be established in several ways.  In the first place, in addition to the usual acoustic modes, there should additionally exist two gapless TO phonon modes that are the `photons' of the emergent  $U(1)$ gauge fields. The speed of sound for these modes vanishes upon approaching any continuous phase boundary to an FE-ordered phase so that a large $T^3$ contribution to the specific heat can be expected. Moreover, deconfined charge excitations corresponding to local violations of the ice rule have creation energy (gap) associated with the energy scale of the ice rule, which is generally much smaller than the electronic gap scale $\sim |\Delta|$ and vanishes upon approaching a continuous transition to the trivial band-insulator phase. These particles have an integer charge with respect to the emergent gauge fields, but an irrational charge, which is a continuous function of $(\lambda - \lambda_c)$, under the EM field~\footnote{In the one-dimensional context, such irrationally charged soliton excitations were investigated theoretically in the classical ($M\to \infty$) limit~\cite{PhysRevLett.49.1455, PhysRevLett.50.439,PhysRevB.28.2653} and more recently in the presence of quantum effects using density-matrix-renormalization group methods~\cite{zhao2024solitons}. That the defects in water ice may carry irrational charge has been noted in Refs.~\cite{PhysRevLett.105.166401, PhysRevB.99.174111}.}. Gapped emergent monopoles should also exist and similarly carry irrational magnetic charges (Sec.~III~B of SM). These electrically or magnetically charged excitations activated at low energy can produce optical conductivity and characteristic noise spectrum~\cite{PhysRevB.97.140402} at temperatures and frequencies well below the electronic band gap.
 
It is also useful to consider identifying features of materials that are somehow `close' to realizing these phases. It would then be sensible to see if various relatively benign perturbations, such as applied pressure or strain, chemical doping, or isotopic substitution, can coax the system across a proximate phase boundary into a pRVB phase.

A trivial insulator close to an ice-rule regime can be expected to exhibit several characteristic features. Most obviously, as a herald of the emergence of the ice rule, it could exhibit a large susceptibility to all forms of ferroelectric orders that preserve the ice rule, indicative of strong collective local quantum fluctuations of dipolar modes. The other side of the same coin is that the corresponding phonon modes (including the TO modes) should become increasingly weakly dispersive and simultaneously soften at all momenta as $\lambda$ approaches $\lambda_c^-$ (Sec.~I~B\&C of SM).

For a material within the ice-rule regime, at temperatures large compared to all the terms that appear in $\hat H^\text{eff}$ but low compared to the energy scale enforcing the ice rule (if such a separation of scales exists), several observable properties can be probed that would indicate the emergence of the ice rule in the patterns of local dipolar distortions. It is well known that the ice rule implies a certain value of residual entropy that depends on the lattice structure (Sec.~I~A of SM). This also results in the existence of ``pinch point(s)'' in correlation functions of the bond dipoles~\cite{PhysRevB.93.125143,PhysRevB.100.014206,PhysRevB.99.174111,henley2010coulomb}. 

Since the quantum term ($J$) essential to any pRVB physics involves the quantum tunneling of ions, one would expect that it will typically be small compared to the potential terms ($V_m$).  This has two consequences for $\lambda > \lambda_c$:  Firstly, it means that the equilibrium state at low $T$ is likely to be one of the topologically trivial broken symmetry states we have mentioned - for instance, a ferroelectric state.  Secondly, all dynamical properties of the dipolar degrees of freedom will be slow, meaning that non-equilibrium states (such as those generic in water ice) may be unavoidable. However, recasting the tunnel action already discussed in a form that is expressed in terms of typical physical parameters, we see that $S \sim z\sqrt{M/m}\ (x_0/a_B)^3$, where $z$ is the number of atoms involved in the resonance ($z=4$ for a square plaquette), $m$ is the electron mass and $a_B $ is the Bohr radius. For a typical value of $\sqrt{M/m}\sim 40 - 600$, this means that $J \sim e^{-S}$ is negligibly small whenever $x_0/a_B$ is order $1$.  However, if through a clever intervention, one can access a regime in which $x_0/a_B$ of order $0.1$ or less, the quantum dynamical terms in $H^\text{eff}$ could be comparable to the potential terms. Thus, a ``trivial'' ferroelectric might be transformable into an FE* or a pRVB. It is also worth noting that more microscopically ``realistic'' versions of related EPC models can be explored numerically as they are free from the fermion-minus-sign problem~\cite{li2019sign,cai2024quantum} and thus may be investigated using quantum Monte Carlo approaches with considerable fidelity (Sec.~VI of SM). 

We would like to conclude with a discussion on the possible variants of the current work for future investigations. Firstly, in two dimensions, the $U(1)$ gauge field will be confined except at the fine-tuned RK point, but `Cantor deconfinement' with incommensurate crystalline orders can arise near this point~\cite{PhysRevB.69.224415,PhysRevB.69.224416}. Secondly, if the band insulator itself is topological, the charge and monopole excitations could mix~\cite{RevModPhys.83.1057,PhysRevB.93.085110}, leading to more exotic responses and surface effects.

{\bf Acknowledgement. } We thank Shivaji Sondhi, Chaitanya Murthy, Vladimir Calvera, Ashvin Vishwanath, Zhi-Qiang Gao, Claudio Castelnovo, Tibor Rakovszky, Jonah Herzog-Arbeitman, Fiona Burnell, Roderich Moessner, Chris Laumann, and Eduardo Fradkin for helpful communications. The work was funded, in part, by NSF grant No. DMR-2310312 at Stanford.

\bibliographystyle{apsrev4-1} 
\bibliography{ref}

\onecolumngrid
\clearpage

\end{document}


\title{Emergent Gauge Fields in Band Insulators: Supplemental Materials}

\author{Zhaoyu~Han}
\affiliation{Department of Physics, Stanford University, Stanford, California 94305, USA} 

\author{Steven~A.~Kivelson} 
\email{kivelson@stanford.edu}
\affiliation{Department of Physics, Stanford University, Stanford, California 94305, USA} 

\maketitle

\onecolumngrid

\tableofcontents

\section{The electron-phonon model}

The model we propose to study can be defined on any lattice in any dimension. Specifically, given a lattice, we will consider its Lieb-type generalization by putting an extra atom orbital on each bond center. The model is aimed to capture the essential effects of the bond atoms' motion in parallel to the bonds' direction. The minimal model then reads:
\begin{align}\label{eq: microscopic model}
    \hat{H} = \sum_{i}\left[\Delta \hat{n}_i - \sum_{j \in \text{nn\ } i, \sigma} t(\hat{X}_{ij})\left(\hat{c}^{\dagger}_{i\sigma} \hat{f}_{\langle ij \rangle  \sigma}+\text{h.c.}\right) \right] + \sum_{\langle ij \rangle} \left(\frac{K\hat{X}_{ij}^2}{2}+\frac{\hat{P}_{ij}^2}{2M}\right) + \dots
\end{align}
where $\hat{c}_{i\sigma}$ and $\hat{f}_{\langle ij \rangle\sigma}$ are the electron annihilation operators vertex $i$ and nearest-neighbor bond $\langle ij\rangle$, respectively. $\Delta$ is the charge transfer gap between vertex and bond orbitals, $\text{nn\ }i$ stands for the nearest neighboring vertices of vertex-$i$, and $\hat{X}_{ij}$ is the displacement of the bond atom relative to the center of bond $\langle ij\rangle$ in the direction of $i$-to-$j$ (so that $X_{ij}=-X_{ji}$ with the origin defined such that  $X_{ij}=0$ when the bond atom is at the bond center). The phonons are assumed to be Einstein phonons with flat bare dispersion $\omega_0\equiv \sqrt{K/M}$, where $K$ is the bare stiffness, and $M$ is the bond atom mass. By writing the model into this form, we have implicitly assumed a (glide) reflection or 180\textdegree-rotation symmetry about each bond center and that all the bonds are equivalent under the space group symmetry of the lattice. The symbol $\dots$ represents the additional realistic terms we will first neglect and treat perturbatively later.

\subsection{Exact ice-rule degeneracy in the static limit $M\rightarrow \infty $}

\label{sec: exact ice rule}

We first study the classical (static phonon) limit of the problem by quenching the phonon dynamics, i.e. taking $M\rightarrow \infty$. We now reveal an exact ice-rule degeneracy for this model in this limit, which holds true as long as the coordination number of every vertex of the lattice is even.

\subsubsection{The degeneracy in of the electronic spectrum}

In the static limit, the ground state energy of the system is determined by the classical phonon configuration $\{X_{ij}\}$. For each phonon configuration, the electrons are described by a non-interacting Hamiltonian, which has a special connectivity structure among the $c$ and $f$ orbitals. Specifically, the single-particle Hamiltonian of the electrons in the basis of $\hat{\psi} = (\hat{c}, \hat{f})$ can be written as:
\begin{align}\label{eq: chiral}
    h(\{X_{ij}\}) \equiv \left[\begin{array}{ c | c }
    \Delta(\{X_{ij}\})  & T(\{X_{ij}\}) \\
    \hline
    T^\dagger(\{X_{ij}\}) & 0
  \end{array}\right] 
\end{align}

The consequence of this structure is that the spectrum of the electrons contains $N^\text{bond}-N^\text{vert.}$ exactly zero modes ($N^\text{vert.}$ and $N^\text{bond}$ are the numbers of vertices and bonds of the lattice, and thus the numbers of $c$- and $f$-orbitals), which are essentially the null modes of matrix $T^\dagger T$ that are distributed on the bond orbitals. The remaining $2N^\text{vert.}$ eigen-energies come in pairs and take the values:
\begin{align}\label{eq: eigenenergy}
    \epsilon_{\pm n} = \frac{\Delta \pm \sqrt{\Delta^2 +4\lambda_n}}{2}
\end{align}
where $\lambda_n$ is the $n$-th eigenstate of a $N^\text{vert.}\times N^\text{vert.}$ matrix, $\Lambda \equiv T T^\dagger$. Since $\Lambda$ is positive semi-definite, $\lambda_n\geq 0$ and thus $\epsilon_{+n}\geq 0$ and $\epsilon_{-n}\leq 0$. We note that $\Lambda$ admits a (single-body) Hamiltonian interpretation since its spectrum is identical to that of the following proxy system defined on the original lattice:
\begin{align}
    \hat{H}_{\text{proxy}} = \sum_{\langle ij \rangle} \Tilde{t}_{\langle ij\rangle} \left(\hat{c}^\dagger_i \hat{c}_j +\text{h.c.}\right) + \sum_i \tilde{\epsilon}_i \hat{n}_i
\end{align}
where $\Tilde{t}_{\langle ij\rangle} \equiv t(\hat{X}_{ij})t(\hat{X}_{ji})$ and $\tilde{\epsilon}_i\equiv \left[\sum_{j\in \text{nn\ } i}t(\hat{X}_{ij})^2\right] $. 
However, $\hat{H}_{\text{proxy}}$ is not really a Hamiltonian since it has units of energy squared.

We now prove a remarkable property of the system resulting from the special structure of the electronic Hamiltonian: Consider the set of phonon configurations in which (1) the magnitude of the atom displacement from the bond-center is the same on every nearest-neighbor bond, i.e.  $|X_{ij}|=x_0$ for all $\langle ij\rangle$, and (2) as many atoms on bonds radiating from each vertex move closer to the vertex as away from it,  i.e.  $\sum_{j\in \text{nn }i} X_{ij}=0$ for any $i$. Although there is an extensive number of such configurations (corresponding to the states of a corresponding vertex model), for given distortion magnitude $x_0$, they all have {\it exactly the same electronic spectrum}. We refer to such configurations as ``ice-rule configurations'' since on the diamond lattice (whose bond lattice is pyrochlore), these correspond precisely to the pattern of H-O displacements identified (without microscopic derivation) by Pauling as the low-energy configurations of water-ice Ic.

The proof is as follows. Property (1) implies $\Tilde{t}_{\langle ij\rangle}= t(x_0)t(-x_0) \equiv \Tilde{t}(x_0)$ is independent of $\langle ij\rangle $, and property (2) further implies that $\tilde{\epsilon}_i = \frac{\mathsf{C}_i}{2} [t(x_0)^2+t(-x_0)^2] \equiv \mathsf{C}_i \Tilde{\epsilon}(x_0)$ is likewise the same for all ice-rule configurations with given $|x_0|$, and depends on the vertex index only through the geometric factor - the coordination number $\mathsf{C}_i$ (number of nearest neighbors) of vertex $i$. These facts together imply that the proxy systems (and thus the $\Lambda$ matrix) for all phonon configurations obeying the generalized ice rule are {\it exactly} the same so that their electronic spectrum is solely determined by $x_0$ but not the further details of the phonon configuration. 

This peculiar property implies a huge degeneracy for phonon configurations obeying the generalized ice rule at the classical level of this problem. Specifically, at any electron filling, for phonon configurations $X_{ij} =x_0 \tau_{ij}$ where $\tau_{ij} =\pm 1$ on every bond and $\sum_{j\in \text{ nn } i} \tau_{ij}$ for any vertex, the classical energy of the system (where electrons occupy the lowest energy modes) only depend on $x_0$:
\begin{align}
    E^{\text{classical}}(\{X^{(I)}_{ij}\}) = E^{\text{ice}}(x_0).
\end{align}

\subsubsection{The ice-rule degeneracy}

There is an extensive number of degenerate ice-rule configurations for fixed $x_0$.  The number can be estimated with Pauling's argument: around a vertex $i$, there are $2^{\mathsf{C}_i}$ or ${\mathsf{C}_i\choose\mathsf{C}_i/2}$ possible $\tau_{ij}$ configurations if they disobey or obey the ice rule, respectively. This gives a `probability' 
\begin{align}
   p_i =\frac{1}{2^{\mathsf{C}_i}} { \mathsf{C}_i \choose \mathsf{C}_i/2 }
\end{align} 
to satisfy the ice rule around vertex $i$. Since the total number of possible $\{\tau_{ij}\}$ configurations disregarding the ice rule is $2^{N^\text{bond}} = 2^{\sum_i \mathsf{C}_i/2 } $, we can estimate the ice-rule degeneracy as
\begin{align}
    g^{\text{ice-rule}} = 2^{\sum_i \mathsf{C}_i/2 } \prod_i p_i = \prod_{i} \left[\frac{1}{2^{\mathsf{C}_i/2}} {\mathsf{C}_i\choose\mathsf{C}_i/2}\right].
\end{align}
For example, the diamond lattice, square lattice, and kagome lattices have the familiar $\ln g^{\text{ice-rule}}  \approx  N^{\text{vert.}} \ln (3/2) \approx 0.18 N^{\text{vert.}}$ whereas the cubic and triangular lattices have $\ln g^{\text{ice-rule}}  \approx  N^{\text{vert.}} \ln (5/2) \approx 0.4 N^{\text{vert.}}$.

\subsubsection{Do the ice-rule configurations minimize the classical energy?}

Here, we prove that the ice-rule configurations are saddle points in the classical energy manifold among configurations with given mean square displacement of phonons, $\overline{X^2} \equiv \sum_{ij} X_{ij}^2 /N^\text{bond} = x_0^2$. The proof holds independent of the value of electron density but requires the aforementioned (glide) reflection or 180\textdegree-rotation symmetry about each bond center and that all the bonds are equivalent under the space group symmetry of the lattice.

For any phonon configuration $\{X_{ij}\}$, the energy gradient can in general be expressed as:
\begin{align}\label{eq: energy gradient}
    \frac{\partial E^\text{classical}}{\partial X_{ij}} = 2 
    \sum_{n,\pm} f_{\pm n}\frac{ \partial \epsilon_{\pm n} }{\partial X_{ij} } =2
    \sum_{n,\pm} \frac{\pm  f_{\pm n}}
    {\sqrt{\Delta^2 +4\lambda_n}} \frac{\partial \lambda_n }{\partial X_{ij} } 
\end{align}
where the summation is over the 
eigenmodes with non-zero energy, $f_{\pm n}=1$ or $0$ depending on whether the mode with energy $\epsilon_{\pm n}$ (defined in Eq.~\ref{eq: eigenenergy}) is occupied, and the factor of 2 accounts for spin degeneracy. Note that the zero modes remain at zero energy regardless of the configuration, so whether or not they are occupied, they do not contribute to the energy gradient. For general configurations, this is a mess. But for an ice-rule configuration $\{X_{ij} = \tau_{ij} x_0 \}$, we find that the gradient has a simple structure:
\begin{align} 
    \frac{\partial E^\text{classical}}{\partial X_{ij}} =\frac{\tau_{ij}}{N^\text{bond}} \left. \frac{\mathrm{d} E^\text{ice}(x)}{\mathrm{d} x}\right|_{x=x_0}
\end{align}
which immediately implies that any small perturbations to the configuration orthogonal to $\tau_{ij}$ and thus $X_{ij}$ cannot change the energy to the first order, so no perturbations that maintain $\overline{X^2} $ change the energy to the first order. 

The proof is as follows. For such a configuration, it can be shown that
\begin{align}\label{eq: H proxy gradient}
    \frac{\partial \hat{H}_\text{proxy}}{\partial X_{ij}} = \tau_{ij}  \sum_{\pm } A_\pm (x_0) \hat{n}_{ij,\pm} + B(x_0) \left(\hat{c}^\dagger_{ij,+} \hat{c}_{ij,-} + \hat{c}^\dagger_{ij,-} \hat{c}_{ij,+}\right)
\end{align}
where [$\dot{t}(x)$ means first derivative of $t(x)$]
\begin{align}
    \hat{c}_{ij,\pm} &\equiv \left(\hat{c}_i \pm \hat{c}_j \right)/\sqrt{2} , \nonumber\\
    \hat{n}_{ij,\pm} &\equiv   \hat{c}^\dagger_{ij,\pm}\hat{c}_{ij,\pm} , \nonumber\\
    A_\pm(x_0) &\equiv \left[\dot{t}(x_0) \mp \dot{t}(-x_0)\right] \left[t(x_0) \pm t(-x_0)\right] ,\nonumber\\
    B(x_0) &\equiv \left[\dot{t}(x_0)t(x_0) +  \dot{t}(-x_0)t(-x_0)\right].
\end{align}
On the other hand, the proxy system has the reflection/rotation symmetry around each bond center $\langle ij\rangle$ inherited from the lattice; therefore, each degenerate subspace of $\hat{H}_\text{proxy}$ form a representation of this symmetry, reducible to the only two one-dimensional irreducible representations corresponding to the two parities under the symmetry transformation. Using this basis, we find that the second term in Eq.~\ref{eq: H proxy gradient} has vanishing contribution to Eq.~\ref{eq: energy gradient} since it is a parity-changing operator. Combined with the fact that all the bonds are equivalent, the energy gradient in Eq.~\ref{eq: energy gradient} is thus proportional to $\tau_{ij}$, with the same prefactor for all bonds.

Note that we have not proven that the ice-rule configurations are the lowest-energy states, i.e. the global minima of the classical energy manifold. We have, however, checked numerically for the electron density $n=2$ (two electrons per unit cell) that this is the case for large clusters (subject to periodic boundary conditions) of a number of regular lattices in one, two, and three dimensions (see Sec.~\ref{sec: cubic numerics} for results of cubic lattice). Indeed, even the value of $x_0$ that minimizes the energy of such configurations can, at present, only be determined numerically.  However, since the many-electron properties of the system in this classical limit are derived from the eigenstates of the non-interacting proxy Hamiltonian, $\hat{H}_\text{proxy}$, it is straightforward to obtain results for infinite systems thanks to the Bloch theorem of periodic lattices. 

For other electron occupation numbers, it is possible that the ice-rule configurations are no longer the classical ground states; for example, one may expect dilute doped holes or electrons to form bipolarons that locally change the optimal configurations.

\subsubsection{The generality of the degeneracy}

The exact ice-rule degeneracy not only holds for the specific Hamiltonian in Eq.~\ref{eq: microscopic model} but pertains in the presence of various kinds of additional terms, such as direct hopping among $c$-orbitals (e.g. $\sum_{\langle ij\rangle}t'(X_{ij}) (\hat{c}^\dagger_{i}\hat{c}_j +\text{h.c.})$), phonon coupling to the local electron densities (i.e. $\sum_{\langle ij\rangle}a(X_{ij}) (\hat{n}_{i}-\hat{n}_j) + b(X_{ij}) (\hat{n}_{i} + \hat{n}_j) + c(X_{ij}) \hat{n}_{\langle ij\rangle} $).  However, many other realistic effects will lift this peculiar degeneracy, such as the direct hopping between $f$ orbitals (e.g. $\sum_{\langle i \langle j\rangle k\rangle } \hat{f}^\dagger_{\langle ij \rangle} \hat{f}_{\langle jk \rangle}$ for pairs of bonds adjacent to a same vertex) and the coupling to other phonon modes (e.g. the motions of vertex atoms). Moreover, at non-zero $\omega_0$, the zero-point motions of the phonons will generically lift this degeneracy among different ice-rule configurations. If these degeneracy-lifting terms are small, we are still left with an effective {\it low-energy} subspace, and the additional terms will become the effective interactions for the phonons.

\subsection{Implication on the phonon spectrum at $\{X_{ij}=0\}$}

\label{sec: implication on phonon spectrum}

In this section, we will investigate the implication of ice-rule degeneracy under the assumption that the ice-rule configurations are the lowest energy states among configurations with given mean square displacement of phonons, $\overline{X^2} \equiv \sum_{ij} X_{ij}^2 /N^\text{bond}$, for small values of $\overline{X^2}$. Again, numerical investigations generally corroborate the validity of this assumption.

The remarkable consequence is that, there are at least $N^\text{bond}-N^\text{vert.}+1$ `flat' phonon modes around the configuration $\{X_{ij}=0\}$, in the sense that these modes have the same eigenfrequency:
\begin{align}
    \omega = \sqrt{\frac{K_{\text{eff}}}{M}}~, \ \ K_{\text{eff} } \equiv \frac{1}{N^{\text{bond}}}\left.\frac{d^2 E^{\text{ice}}(x)}{d x ^2}\right|_{x=0}.
\end{align}

The proof is as follows. First, for any phonon fluctuation obeying the ice rule, we have 
\begin{align}
    &E^{\text{classical}}(\{X_{ij}= \delta x \cdot  \tau_{ij}\}) \approx \frac{\delta x^2}{2} \sum_{\langle ij\rangle, \langle kl\rangle} \tau_{ij} K_{ij,kl} \tau_{kl}  \nonumber\\
    =&E^{\text{ice}} (\delta x) \approx \frac{1}{2}  K_{\text{eff}} N^{\text{bond}} \delta x^2
\end{align}
where 
\begin{align}
    K_{ij,kl} \equiv \left. \frac{\partial^2 E^{\text{classical}}}{\partial X_{ij}\partial X_{kl}} \right|_{X=0}
\end{align}
is the stability matrix of the system. Therefore, 
\begin{align}
   K_{\text{eff}} =  \sum_{\langle ij\rangle, \langle kl\rangle} \frac{ \tau_{ij}}{N^{\text{bond}}} K_{ij,kl} \frac{ \tau_{kl}}{N^{\text{bond}}}
\end{align}
 for any ice rule configurations. Since we have assumed that the ice-rule configurations have the smallest energy when compared with other configurations with the same $\overline{X^2}$, the above equality implies that each ice-rule mode defined by eigenvector $\tau_{ij}/\sqrt{N^{\text{bond}}}$ is an eigenmode of the system with stiffness $K_{\text{eff}}$.

Then, any linear combinations of the ice-rule modes will also be an eigenmode with the same stiffness and, thus, frequency. To count how many linear independent modes there are, we recognize the fact that differentiating two ice-rule modes will always generate a certain `loop' mode, i.e. $\delta \tau_{ij} $ is only uniformly non-zero along oriented loops. The number of linearly independent ice-rule modes thus is lower bounded by that of linearly independent loops in the system. We note that any lattice has two types of loops: contractible loops, which are generated by linear combining the smallest loops - plaquettes, and non-contractible loops, which depend on the topology of the underlying manifold. However, simply adding up those numbers will overestimate the number of linearly independent loops since if several plaquettes (faces) enclose a polyhedron, linear combining the corresponding loops will vanish. We need to account for those constraints, each given by a 3D object in the system. However, again, those 3D constraints are not linearly independent if some of them further enclose 4D objects, etc. After properly counting the number of constraints, we reach an expression for the number of linearly independent loops (and thus a lower bound on the number of flat phonon modes)
\begin{align}
    N^{\text{flat phonon modes}} \ge  (N^{2-\text{cell}} + b_1) -  (N^{3-\text{cell}} + b_2) + (N^{4-\text{cell}} + b_3)  \dots
\end{align}
where $N^{k-\text{cell}}$ is the number of $k$-dimensional cells (e.g. $0$-cells are vertices, $1$-cells are bonds, $2$-cells are plaquettes, $3$-cells are polyhedra) in the system, $b_k$ is the $k$-th Betti number representing the number of $k$-dimensional non-contractible cycles. Using the different definitions of Euler characteristic:
\begin{align}
    \chi = \sum_{k=0}^{\infty} (-1)^k N^{k-\text{cell}} = \sum_{k=0}^{\infty} (-1)^k b_k.
\end{align}
we reach the conclusion that
\begin{align}
    N^{\text{flat phonon modes}} \ge N^{\text{bond}}-N^{\text{vert.}} + b_0
\end{align}
where $b_0$ counts the number of disconnected clusters ($=1$ in the context we are considering).

\subsection{Numerical results for cubic lattice at $n=2$ in the static limit}
\label{sec: cubic numerics}

With the above general property in mind, from here on, we will focus on three-dimensional cubic lattices with electron filling $n=2$, i.e., two electrons per unit cell. We will set the lattice constant $a_0=1$ unless otherwise stated. To obtain more results through numerical simulations, we take the leading order expansion of
\begin{align}
    t(x) \approx t_0 - g x
\end{align}
which is well justified as long as the phonon displacements are sufficiently small. Now we note that there are three independent energy scales in the static limit of the problem: $\Delta$, $t_0$ and $\Ueph \equiv g^2/K $ (a characteristic energy scale for the electron-phonon coupling strength). Therefore, two dimensionless parameters, $\lambda \equiv \Ueph/t_0$ and $\xi \equiv \Delta/t_0$, can fully characterize the problem.

Then, we performed numerical optimization (with a gradient descent algorithm) over all possible phonon configurations in a finite-size cluster of size up to $9\times 9\times 9$. We have verified that, for various sets of parameters we examined, the classical groundstate configurations always obey ice rules (including the trivial possibility $x_0=0$).

\begin{figure}
    \subfigure[]{\includegraphics[width = 0.32\linewidth]{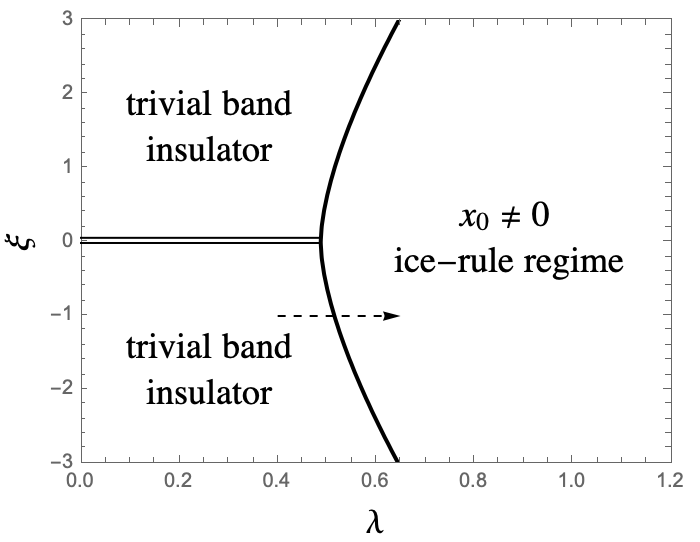} \label{fig: static phase diagram}}
    \subfigure[]{\includegraphics[width = 0.32\linewidth]{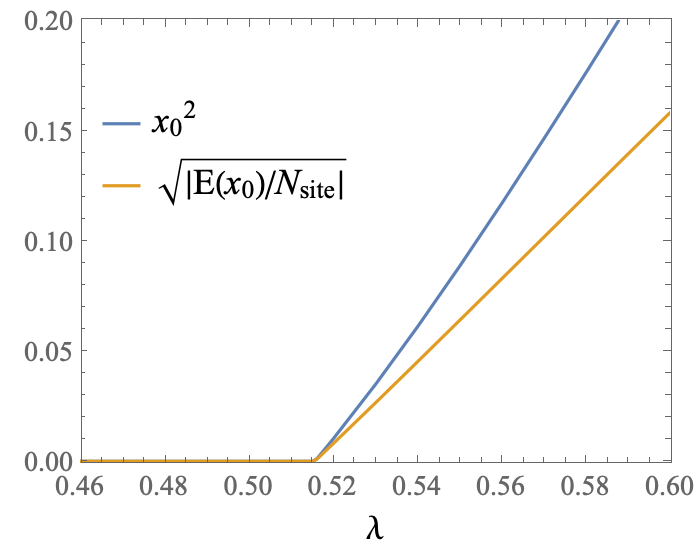} \label{fig: x0 energygain}}
    \subfigure[]{\includegraphics[width = 0.32\linewidth]{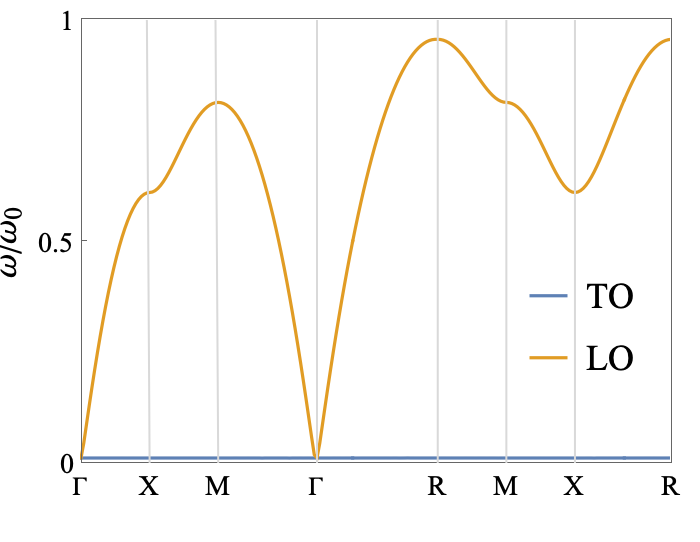} \label{fig: phonon spectrum}}

    \caption{The numerical results for the electron-phonon model defined in Eq.~\ref{eq: microscopic model} on the cubic lattice at electron density $n=2$ in the static limit. (a) The phase diagram. The axes represent dimensionless parameters $\lambda \equiv \Ueph/t_0$ and $\xi \equiv \Delta/t_0$. The solid line is the phonon transition boundary, and the double line represents an electronic transition between two band insulators, on which the system is a (fine-tuned) Dirac semimetal with two extra flat bands at the Fermi level. (b) Along the cut indicated by the dashed arrow in the phase diagram ($\xi =-1$), the optimal $x_0$ and the corresponding energy gain per unit cell. In the plot, the length unit is $\sqrt{t_0/K}$, and the energy unit is $t_0$. (c) The phonon spectra at $\xi=-1$ and the corresponding critical $\lambda_c\approx 0.5154$. TO stands for transverse optical mode, and LO stands for longitudinal optical mode. }
    
\end{figure}

With this confirmation, we further write down the energy of such configurations explicitly:
\begin{align}
    E^{\text{ice}}(x_0)/N^{\text{vert.}} = \frac{3 Kx_0^2}{2}+ 2 \iiint_{0}^{2\pi} \frac{d^3\bm{k}}{(2\pi)^3} \epsilon_-(\bm{k})
\end{align}
where 
\begin{align}
    \epsilon_-(\bm{k}) = \frac{\Delta-\sqrt{\Delta^2+ 8(t_0^2-g^2x_0^2)(\cos k_x + \cos k_y + \cos k_z)+24(t_0^2+g^2 x_0^2)}}{2}
\end{align}
is the dispersion of the lowest band. This energy expression can be evaluated with arbitrary accuracy. By examining the optimal $x_0$ for $E^{\text{ice}}(x_0)$, we obtained a phase diagram on the plane of dimensionless parameters $\lambda$ and $\xi$ in Fig.~\ref{fig: static phase diagram}. We find that, for any values of $\Delta$, as long as $\Ueph$ surpasses certain critical value (which is comparable to $\Delta$ and $t_0$), ice-rule configurations with non-zero $x_0$ will be favored. We emphasize that when either $\Delta$ or $x_0$ is nonzero, the electronic sector is a band insulator with a finite gap between the bottom of the higher dispersing bands and the top of the lower dispersing band in the electronic spectrum
\begin{align}
    \epsilon_{\text{gap}} = \sqrt{\Delta^2+48g^2 x_0^2}.
\end{align}

It is hard to analytically determine the value of critical $\lambda_c$ for the generical values of $\xi$. But when $|\xi|\gg 1$, it is straightforward to expand $\epsilon_-(\bm{k})$ in powers of $1/\xi$ and $x_0$, and thus obtain $\lambda_c \approx |\xi|/8$ in this limit, i.e. the critical $U_{\text{e-ph},c} \approx |\Delta|/8$ when $|\Delta|\gg t_0$.

To further investigate the nature of the transition, we zoom in to the neighborhood of a transition point corresponding to $\xi=-1$ at $\lambda_c \approx 0.5154$. The results are shown in Fig.~\ref{fig: x0 energygain}. We find that the $x_0$ continuously develops from the transition point as
\begin{align}
    x_0 \propto \sqrt{\lambda-\lambda_c} \ , \ \ \lambda>\lambda_c 
\end{align}
and the corresponding energy gain develops as $(\lambda-\lambda_c)^2$. Both facts are consistent with the Landau theory of phase transitions, which assumes the general expansion of the energy
\begin{align}\label{eq: expansion}
    E^{\text{ice}}(x_0)/N^{\text{vert.}} \approx - (\lambda-\lambda_c) a x_0^2 + b x_0^4
\end{align}
near the critical point.

Lastly, we numerically examine the sharp prediction about the number of flat phonon modes revealed in the previous section. This can be done straightforwardly with second-order perturbation theory by treating the small phonon vibrations as perturbations of the electron Hamiltonian. The resulting phonon spectrum $\omega(\bm{q};\xi=-1,\lambda=\lambda_c)$ at the transition point is plotted in Fig.~\ref{fig: phonon spectrum}. [We note that the phonon spectrum at other $\lambda<\lambda_c$ can be obtained by recognizing that $[\omega^2(\bm{q};\xi=-1,\lambda)-1]/\lambda$ is independent of $\lambda$.] We find that the two transverse (TO) modes whose polarization vectors $\bm{l}(\bm{q})$ satisfy 
\begin{align}
    \bm{l}(\bm{q})\cdot(\sin k_x, \sin k_y, \sin k_z) = 0
\end{align}
are exactly flat (independent of $\bm{q}$). The frequency, $\omega$, is degenerate with the longitudinal (LO) mode at $\bm{q}=(0,0,0)$. Therefore, there are $2N^\text{vert.} + 1 = N^\text{bond} - N^\text{vert.} +1 $ flat phonon modes in the system (assuming periodic boundary conditions in all directions by putting the lattice on a 3-torus), in agreement with our result in Sec.~\ref{sec: implication on phonon spectrum}.

\subsection{Scaling of the emergent scales near the transition point}

\label{sec: scaling}

Within the ice-rule regime, there are several emergent scales, whose scaling behavior near the transition point (at a fixed $\xi\sim \mathcal{O}(1)$) can be obtained by using the expansion in Eq.~\ref{eq: expansion}:
\begin{itemize}
    
    \item {\bf The creation energy of a topological defect, $E^{\text{defect}}$.} When the ice rule is violated around a lattice vertex, e.g. by having a 4-in-2-out instead of 3-in-3-out configuration, a topological defect is created. Such a topological defect will result in an energy loss in a few unit cells around the lattice vertex. Therefore, this energy scale can be estimated as 
    \begin{align}
        E^{\text{defect}}\sim (\lambda-\lambda_c)^2 t_0.
    \end{align}

    \item {\bf The renormalized TO phonon frequency, $\omega$.} The phonon frequencies for different $\{X_{ij}\}$ configurations are generically different; however, for those modes that are orthogonal to the ice rule constraint, their frequencies are of the order
    \begin{align}
    \omega \sim \sqrt{\frac{1}{N^{\text{bond}}M}\left.\frac{d^2 E^{\text{ice}}(x)}{d x ^2}\right|_{x=x_0}} \sim \sqrt{\lambda-\lambda_c}\omega_0
    \end{align}
    This also estimates the difference in the zero point energy (ZPE) between different ice-rule configurations. 
    
    \item {\bf The tunneling amplitude between two ice-rule configurations through a minimal flip on a plaquette, $J$.} When the system is not at static limit (i.e. $\omega_0 \neq 0$), the phonon fluctuation will cause quantum tunneling between different classically degenerate configurations through instanton events. This amplitude generically takes the form
    \begin{align}
        J \sim \omega \sqrt{S}\mathrm{e}^{-S}
    \end{align}
    where $S$ is the action of the instanton tunneling event:
    \begin{align}
        S = \int^{\{X\}_{\text{final}}}_{\{X\}_\text{initial}} D\{X\} \sqrt{2M \left[E(\{X\})-E_\text{GS}\right]}.
    \end{align}
    This action can be estimated as 
    \begin{align}
        S \sim x_0 \sqrt{M E^{\text{defect}}} \sim (\lambda-\lambda_c)^{3/2} t_0/\omega_0.
    \end{align}
    This leads to
    \begin{align}
        J \sim \sqrt{t_0 \omega_0} (\lambda-\lambda_c)^{5/4} \exp\left[-c (\lambda-\lambda_c)^{3/2} \frac{t_0}{\omega_0}\right]
    \end{align}
    with $c$ an $\mathcal{O}(1)$ number. Note that, for this estimation to be justified in the low-energy subspace (i.e. the barriers between ice-rule configurations are high enough), we need to have $S\gg 1$.

    \item {\bf The electric dipole moment $\eta$ carried by each bond. } When the atom on a bond $ij$ is displaced, the physical charge on the atom $Q$ is also displaced, resulting a net dipole $X_{ij} Q$ in the $i$-to-$j$ direction. Since $Q$ should, in general, be an $\mathcal{O}(1)$ number, the magnitude of the dipole can be estimated as $\eta \sim x_0$. 
    
\end{itemize}

\section{The effective model}

Within the degenerate subspace consisting of ice-rule configurations, we are motivated to write down an effective model for the dipolar phonon degree of freedom on each bond. Since all $X_{ij}$ have the same amplitude $x_0$, they can be parametrized by a dipole variable $\tau_{ij}$ on each bond
\begin{align}
X_{ij} = \tau_{ij} x_0.
\end{align}
Pictorially, these dipole variables can be represented by arrows on bonds. The ice rule becomes a constraint on the Hilbert space of the problem by restricting each vertex to only admit 3-in-3-out arrow configurations. Now, we aim to propose an effective model within this low-energy subspace.

Before we introduce the model, we first note that each square plaquette (pictorially represented by $\plaquette$) can take four types of edge configurations. The representative configurations of each type (which are related to other equivalent configurations by rotation and mirror reflection) are, respectively, type 0: \plqzero, type 1:  \plqone, type 2: \plqtwo, and type 3: \plqthree. We denote the projector onto the subspace corresponding to type-$m$ configurations on a plaquette as $\hat{P}_m^{\smallplaquette}$.

The model then reads:
\begin{align}\label{eq: model}
\hat{H} =  \sum_{\smallplaquette} \left[\sum_{m=1,2,3} V_m  \hat{P}_m^{\smallplaquette} - J \left(\left|\plqthree \right\rangle \left\langle \plqthreereverse \right| + \text{h.c.}\right)\right]
\end{align}
where we have set the energy of type-0 plaquettes to zero, i.e. $V_0\equiv 0$, without loss of generality (since the numbers of different types of plaquettes add up to a constant, $3N^\text{vert.}$). 
Since $J$ is a positive quantum tunneling amplitude, we can set it as the energy unit. Therefore, there are only three independent dimensionless parameters in this problem: 
\begin{align}
    v_{m=1,2,3} \equiv V_{m=1,2,3}/J.
\end{align}

This is a variant of the quantum vertex model in the sense that the classical interaction terms depend on not only the vertices status but also the plaquettes status (See Sec.~\ref{sec: vertex constraints} for why we say vertex status has been taken into account in this formulation). One can, of course, include longer-ranged interactions and tunnelings (involving more bonds in more plaquettes) into this model, but we neglect these terms under the assumption that those couplings decay rapidly as the size of the cluster considered increases.

\subsection{Topological sectors and Extensive conserved quantities}

The ice rule constraint on the Hilbert space immediately implies that the allowed arrow configurations can be classified into different topological sectors. Each sector $\mathcal{T}$ can be labeled by the total flux numbers $(\mathcal{T}_x, \mathcal{T}_y, \mathcal{T}_z)$ in the three directions, which are defined as:
\begin{align}
    \mathcal{T}_{\alpha=x,y,z} = \sum_{\bm{r} \in S_{\bm{r}_{a}=\bm{r}_{a,0}} } u_{\bm{r},\bm{r}+\bm{\alpha}}.
\end{align}
where $S_{\bm{r}_{\alpha}=\bm{r}_{\alpha,0}}$ is a spatial slice corresponding to an arbitrary fixed $\bm{r}_{\alpha}=\bm{r}_{\alpha,0}$. $\bm{\alpha}$ is the unit vector in $\alpha$-direction. [The ice rule ensures that $\mathcal{T}_{\alpha}$ does not depend on the choice of $\bm{r}_{\alpha,0}$.] Different configurations in different sectors are not connected by the Hamiltonian since flipping the dipoles along any contractible loop does not change the value of $\mathcal{T}_{\alpha}$.

Related to the above observations, we identify an extensive number of conserved quantities of this problem. Viewing $\hat{\tau}_{ij}$ as an operator and defining $\hat{\tau}_{\alpha}(\bm{k})\equiv \sum_{i} \mathrm{e}^{\mathrm{i} \bm{r}_i \cdot \bm{k}} \hat{\tau}_{\bm{r}_i,\bm{r}_i+\bm{\alpha}} /\sqrt{N^\text{vert.}}$, we have
\begin{align} \label{eq: conserved quantities}
    [\hat{\tau}_{a}(\bm{k}) , \hat{H}] =0 \text{ if } \bm{k}\propto \bm{\alpha}
\end{align}
This is because flipping dipoles on any plaquette does not change the value of $\sum_{\bm{r}\in S_{\bm{r}_{\alpha}=\bm{r}_{\alpha,0}}} \hat{\tau}_{\bm{r},\bm{r}+\bm{\alpha}}$ for any slice $\bm{r}_{\alpha,0}$, and $\hat{\tau}_\alpha(\bm{k})$ with $\bm{k}\propto \bm{\alpha}$ is a linear combination of such sums at different $\bm{r}_{\alpha,0}$. 

Putting in modern language, these conserved quantities are the conserved symmetry charges of the 1-form electric symmetry of the system~\cite{tong2018gauge,thorngren2023higgs}.

\subsection{Vertex interactions and vertex constraints}

\label{sec: vertex constraints}

Note that the model in Eq.~\ref{eq: model} only contains plaquette terms. One may wonder why the vertex terms are not explicitly included here, since there can definitely be important short-distance interactions that are determined by the arrow configuration around each vertex (pictorially represented by \vertex). To explain this, we first sort the different arrow configurations around a vertex into two types, which we call coplanar and non-coplanar types, depending on whether the in-going arrows are coplanar or not. The representative configurations of each type (which are related to other equivalent configurations by rotation and mirror reflection) are respectively coplanar: \vertexcoplanar, and non-coplanar:  \vertexnoncoplanar. Then we recognize a set of geometric constraints (originating from the generalized ice rule) relating the number of different types of plaquettes and vertices
\begin{align}
&2N^\text{vert.}_\text{coplanar} + 3N^\text{vert.}_\text{non-coplanar} = N^{\text{corner}}_{\text{crowded/deserted}} = N_0^\text{plaq.} + 2N_1^\text{plaq.} + N_2^\text{plaq.} \\
&8N^\text{vert.}_\text{coplanar} + 6N^\text{vert.}_\text{non-coplanar} = N^{\text{corner}}_{\text{normal}} = 2N_0^\text{plaq.} + 2N_2^\text{plaq.} + 4N_3^\text{plaq.}
\end{align}
which can be obtained by counting different types of ``corners'' of plaquettes from two different perspectives. Specifically, if two/one/zero arrows run into a corner of a plaquette, we call the corner ``crowded''/``normal''/``deserted'' respectively. Linear recombining the equations, we obtain:
\begin{align}
N^\text{vert.}_\text{coplanar} &= - N_1^\text{plaq.} +  N_3^\text{plaq.} \\
    N^\text{vert.}_\text{non-coplanar} &= \frac{1}{3} \left(N_0^\text{plaq.} + 4 N_1^\text{plaq.} + N_2^\text{plaq.} - 2 N_3^\text{plaq.} \right)
\end{align}
Therefore, even if there are interaction terms associated with the vertex configurations, they can be converted into interaction terms that only depend on plaquette configurations.

Furthermore, realizing that $0\leq N^{\text{vert.}}_{\text{coplanar}}, N^\text{vert.}_\text{non-coplanar} \leq N^\text{vert.}$, we obtain a non-trivial constraint on the numbers of type-1 and type-3 plaquettes:
\begin{align}
   N_3^\text{plaq.} - N^\text{vert.} \leq N_1^\text{plaq.} \leq N_3^\text{plaq.}
\end{align}
When the first (second) equal sign is taken, all the vertices are coplanar (non-coplanar).

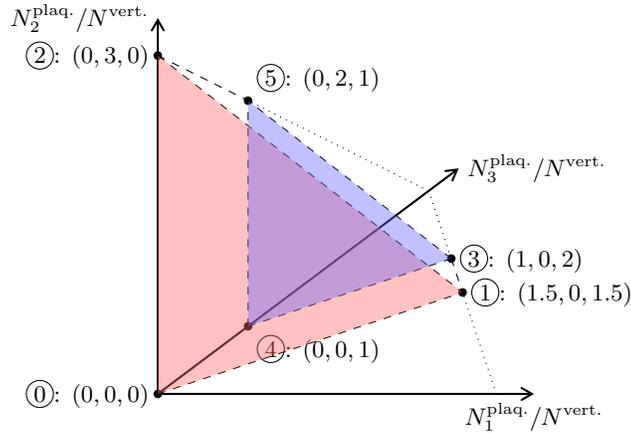
\begin{figure}
    \centering
    \begin{tikzpicture}
\draw coordinate (direction one) at (5,0);
\draw (5,0) node[anchor=north] {$N^{\text{plaq.}}_1/N^\text{vert.}$};
\draw coordinate (direction two) at (0,5);
\draw (0,5) node[anchor=east] {$N^{\text{plaq.}}_2/N^\text{vert.}$};
\draw coordinate (direction three) at (4,3);
\draw (4,3) node[anchor=west] {$N^{\text{plaq.}}_3/N^\text{vert.}$};
\draw coordinate (zero) at (0,0); \filldraw [black] (zero) circle (0.05) node [anchor=east]{\circled{0}: $(0,0,0)$};
\draw coordinate (one) at (4.05,1.35); \filldraw [black] (one) circle (0.05) node [anchor=west]{\circled{1}: $(1.5,0,1.5)$};
\draw coordinate (two) at (0,4.5); \filldraw [black] (two) circle (0.05) node [anchor=east]{\circled{2}: $(0,3,0)$};
\draw coordinate (three) at (3.9,1.8); \filldraw [black] (three) circle (0.05) node [anchor=west]{\circled{3}: $(1,0,2)$};
\draw coordinate (four) at (1.2,0.9); \filldraw [black] (four) circle (0.05) node [anchor=north west]{\circled{4}: $(0,0,1)$};
\draw coordinate (five) at (1.2,3.9); \filldraw [black] (five) circle (0.05) node [anchor=south west]{\circled{5}: $(0,2,1)$};
\draw coordinate (fictitious three) at (3.6,2.7);
\draw coordinate (fictitious one) at (4.5,0);
    \draw[-angle 60,thick] (zero) -- (direction one);
    \draw[-angle 60,thick] (zero) -- (direction two);
    \draw[-angle 60,thick] (zero) -- (direction three);
    \draw[dashed, thin] (zero) -- (one);
    \draw[dashed, thin] (three) -- (one);
    \draw[dashed, thin] (four) -- (three);
    \draw[dashed, thin] (four) -- (five);
    \draw[dashed, thin] (two) -- (five);
    \draw[dashed, thin] (two) -- (one);
    \draw[dashed, thin] (three) -- (five);
    \draw[dotted, thin] (fictitious three) -- (three);
    \draw[dotted, thin] (fictitious one) -- (one);
    \draw[dotted, thin] (fictitious three) -- (five);
    \fill[red!60, fill opacity=0.4] (zero) -- (one) -- (two) -- (zero);
    \fill[blue!60, fill opacity=0.4] (four) -- (three) -- (five) -- (four);
\end{tikzpicture}
    \caption{An illustration of allowed values of $N_{i}^{\text{plaq.}}$. The blue and red rectangular surfaces represent the coplanar and non-coplanar limits. }
    \label{fig: allowed N plaquette}
\end{figure}

Incorporating the above non-trivial constraints on the numbers of plaquettes of different types $N^{\text{plaq.}}_{m=0,1,2,3}\ge 0$, and recognizing that $\sum_{m=0,1,2,3} N^{\text{plaq.}}_m = 3N^{\text{vert.}}$, we find that the allowed values of $\{N^{\text{plaq.}}_1,N^{\text{plaq.}}_2,N^{\text{plaq.}}_3\}$ can be represented by the region between the colored surfaces in Fig.~\ref{fig: allowed N plaquette}.

\subsection{Classical limits}

When $J$ is small compared to $V$'s (in the sense that will be made precise below), its effects can be neglected, and the problem becomes classical. One of the vertices \circled{0}\dots \circled{5} in Fig.~\ref{fig: allowed N plaquette} will have the lowest energy as a result of optimization on this convex polygon. These are the classical limits of the effective model in Eq.~\ref{eq: model}. In Table~\ref{table: classical limits}, we illustrate (within a $2\times 2\times 2$ supercell) all the states we found in each limit and the corresponding parametric limit. We obtained a rich host of ferroelectric, anti-ferroelectric, and their coexisting (ferri-electric) phases. These states considered here all have the feature that the ordering wavevectors of all the polarization directions of the phonons, $\Vec{Q}^{(\alpha=x,y,z)}$, contain only $0$ or $\pi$ in their components. We note that some of the states have analogs in the study of octahedral spin ices~\cite{PhysRevLett.129.247201,PhysRevB.88.024407,pickles2008critical,PhysRevB.71.125102}.

\begin{table}[H]
    \centering
    \begin{tabular}{|c|c|c|c|c|}
    \hline
      Name   & Parametric limit  & Representative $[\Vec{Q}^{(\alpha)}]_\beta$  & Degeneracy & Illustration \\
        \hline
       \circled{0} & $v_1+v_3,v_2,v_3\gg 1$ & $\begin{bmatrix}
          ( 0 & 0 & 0 )\\
          ( 0 & 0 & 0 )\\
          ( 0 & 0 & 0 )
       \end{bmatrix}_{\alpha\beta}$ & $2^3 \times 1=8$ & \configzero\\
         \hline
       \circled{1} & $\begin{aligned}
            v_3-v_1,-v_1-v_3, \\
            2v_2-v_1-v_3 \gg 1
        \end{aligned}$  & $\begin{bmatrix}
           ( 0 & \pi & \pi ) \\
           ( \pi & 0 & \pi ) \\
           ( \pi & \pi & 0 )
       \end{bmatrix}_{\alpha\beta}$ & $2^3\times 1=8$ & \configone \\
         \hline
        \circled{2}a &  \multirow{2}{*}[-2em]{$\begin{aligned}
            &v_3-v_2,-v_2, \\
            &v_1+v_3-2v_2 \gg 1
        \end{aligned}$}  &  $\begin{bmatrix}
           ( 0 & \pi & \pi ) \\
           ( 0 & 0 & \pi ) \\
           ( 0 & 0 & 0 )
       \end{bmatrix}_{\alpha\beta}$ &  $2^3\times 6=48$ & \configtwoa \\
         \cline{1-1}\cline{3-5}
        \circled{2}b &  & $\begin{bmatrix}
           ( 0 & \pi & 0 ) \\
           ( 0 & 0 & \pi ) \\
           ( \pi & 0 & 0 )
       \end{bmatrix}_{\alpha\beta}$  & $2^3\times 2 = 16$  & \configtwob\\
         \hline
         \circled{3} & $\begin{aligned}
            &v_1-v_3,-v_1-v_3, \\
            &2v_2-v_1-v_3 \gg 1
        \end{aligned}$  & $\begin{bmatrix}
           ( \pi & \pi & \pi ) \\
           ( \pi & \pi & \pi ) \\
           ( \pi & \pi & 0 )
       \end{bmatrix}_{\alpha\beta}$ &  $2^2\times 3 = 12$ & \configthree\\
         \hline
        \circled{4} & $v_1+v_3,v_2,-v_3\gg 1$ & $\begin{bmatrix}
           ( \pi & \pi & 0 ) \\
           ( \pi & \pi & 0 ) \\
           ( 0 & 0 & 0 ) 
       \end{bmatrix}_{\alpha\beta}$  & $2^2\times 3 = 12$ & \configfour\\
         \hline
        \circled{5}a & \multirow{3}{*}[-4em]{$\begin{aligned}
            &v_2-v_3,-v_2, \\
            &v_1+v_3-2v_2 \gg 1
        \end{aligned}$} & $\begin{bmatrix}
           ( \pi & \pi & \pi ) \\
           ( \pi & \pi & \pi ) \\
           ( 0 & 0 & 0 ) 
       \end{bmatrix}_{\alpha\beta}$  & $2^2\times 3 = 12$ & \configfivea\\
         \cline{1-1}\cline{3-5}
        \circled{5}b   &   & $\begin{bmatrix}
           ( \pi & \pi & \pi ) \\
           ( \pi & \pi & 0 ) \\
           ( 0 & \pi & 0 ) 
       \end{bmatrix}_{\alpha\beta}$ & $2^2\times 6 =24$ & \configfiveb \\
         \cline{1-1}\cline{3-5}
        \circled{5}c &  & $\begin{bmatrix}
           ( \pi & \pi & 0 ) \\
           ( \pi & \pi & 0 ) \\
           ( \pi & \pi & 0 )
       \end{bmatrix}_{\alpha\beta}$ & $2^2\times 3=12$ & \configfivec \\
         \hline
    \end{tabular}
    \caption{The classical limits of the model in Eq.~\ref{eq: model}. Arrows (representing $\tau_{ij}$) in {\color{red} $x$ (red)}, {\color{black!40!green} $y$ (green)}, {\color{blue} $z$ (blue)} directions are colored differently.   }
    \label{table: classical limits}
\end{table}

\subsection{A solvable point with RVB groundstates}

When $v_1=v_2 =0$ and $ v_3=1$ , the Hamiltonian can be reorganized into:
\begin{align}\label{eq: RK point model}
\hat{H} =  J \sum_{\smallplaquette}  \left(\left|\plqthree \right\rangle -  \left| \plqthreereverse \right\rangle \right) \left(\left\langle \plqthree \right| - \left\langle \plqthreereverse \right| \right) 
\end{align}
Therefore, the ground state can be exactly obtained by equally superposing all the configurations $C$ connected by the resonance within each topological sector $\mathcal{T}$:
\begin{align}
    |\Psi_\mathcal{T}\rangle \propto \sum_{C\in \mathcal{T}} |C\rangle
\end{align}
Those states are the ground states since the Hamiltonian Eq.~\ref{eq: RK point model} is positive semi-definite, and these states are annihilated by each term in the Hamiltonian and thus have exactly zero energy. Rokhsar and Kivelson (RK) first recognized this kind of solvable point in a close cousin of the current model - the quantum dimer model~\cite{PhysRevLett.61.2376}. This point is thus called the RK point. This point has a host of important properties, which we will discuss below. This state is literally a ``resonating valence bond'' (RVB) state, and as we will see below, it is a transition point out of an RVB phase. Since the underlying active degrees of freedom are phonons, we call this phase ``phononic RVB'' (pRVB).

\subsubsection{Correlation functions}
Due to the equal superposition, the correlation functions in these groundstate wavefunctions are equal to that in the classical arrowed ice-rule vertex model problem at infinite temperature. The dipole-dipole correlation functions can be evaluated with Monte Carlo~\cite{PhysRevB.69.064404,balasubramanian2022exact,PhysRevLett.91.167004}, which has the asymptotic form ($\alpha,\beta = x,y,z$)
\begin{align}\label{eq: classical correlation function}
    \langle \hat{\tau}_{\bm{r},\bm{r}+\bm{\alpha}} \hat{\tau}_{\bm{r}',\bm{r}'+\bm{\beta}} \rangle \rightarrow \frac{1}{4\pi \kappa_\tau }  \frac{1}{|\delta \bm{r}|^5} \left(|\delta \bm{r}|^2 \delta_{\alpha \beta} - 3\delta \bm{r}_\alpha \delta \bm{r}_\beta \right) \text{ as } |\delta \bm{r}|\rightarrow \infty
\end{align}
with $\delta\bm{r} = \bm{r}-\bm{r}'$. $\kappa_\tau$ is a constant fully determined by the cubic lattice structure, whose precise value is unfortunately unknown but can be estimated to be $\kappa_\tau \approx 0.9$ according to the numerical data in Ref.~\cite{PhysRevB.69.064404}. This amounts to a structure factor
\begin{align}\label{eq: classical structure factor}
    S_{\alpha \beta}(\bm{k}) = \langle \hat{\tau}_{\alpha} (-\bm{k}) \hat{\tau}_{\beta}(\bm{k})\rangle = \frac{1}{\kappa_\tau} \left(\delta_{\alpha \beta} - \frac{\bm{k}_\alpha \bm{k}_\beta}{|\bm{k}|^2}\right) \text{ as } \bm{k} \rightarrow \bm{0}
\end{align}
where $\hat{\tau}_{\alpha}(\bm{k})\equiv \sum_{i} \mathrm{e}^{\mathrm{i} \bm{r}_i \cdot \bm{k}} \hat{\tau}_{\bm{r}_i,\bm{r}_i+\bm{\alpha}}/\sqrt{N^\text{vert.}}$.

\subsubsection{Transverse gapless modes}

Moreover, by adopting single-mode approximation (SMA), one can confirm the existence of two transverse, gapless modes dispersing as $|\bm{k}|^2$~\cite{PhysRevB.68.184512,PhysRevLett.61.2376, moessner2010quantum}. The proof is as follows. For any $\bm{k}$, we may construct a trial wavefunction with polarization vector $\bm{l}$
\begin{align}
    |\bm{l},\bm{k}\rangle \equiv \sum_{\alpha=x,y,z} \bm{l}_\alpha \hat{\tau}_{\alpha}(\bm{k}) |GS\rangle 
\end{align}
The energy of this state, which upper bounds the lowest excitation energy, is 
\begin{align}
    E_\text{trial}(\bm{l},\bm{k}) = \langle \bm{l},\bm{k} |\hat{H}|\bm{l},\bm{k}\rangle / \langle \bm{l},\bm{k} |\bm{l},\bm{k}\rangle 
\end{align}
The denominator can be evaluated to $\sim |\bm{l}|^2 - |\bm{l}\cdot \bm{k}|^2/|\bm{k}|^2 = |\bm{l}\times \bm{k}|^2 /|\bm{k}|^2$, whereas the numerator can be calculated as (remember $\hat{H}|GS\rangle =0$)
\begin{align}
    \langle \bm{l},\bm{k} |\hat{H}|\bm{l},\bm{k}\rangle = \sum_{\alpha \beta} \bm{l}_\alpha \bm{l}_\beta \langle GS | \left[ \hat{\tau}_{\alpha}(-\bm{k}), \left[\hat{H}, \hat{\tau}_{\beta}(\bm{k}) \right] \right] |GS\rangle.
\end{align}
The conserved quantities in Eq.~\ref{eq: conserved quantities} imply that the numerator cannot depend on $\bm{k}_\parallel \equiv \bm{l} (\bm{l}\cdot \bm{k}) /|\bm{l}|^2$, but can only depend on $\bm{k}_\perp \equiv \bm{k}-\bm{k}_\parallel$ at small $\bm{k}$. The inversion symmetry of the system and the requirement of analyticity of $E_\text{trial}(\bm{l},\bm{k})$ further restrict that the numerator can grow as fastest as $|\bm{k}_\perp|^2 =  |\bm{l} \times \bm{k}|^2/|\bm{l}|^2$. Putting everything together, we reach the conclusion that the existence of two linearly independent states with two possible polarizations $\bm{l}\perp \bm{k}$ proves the existence of two transverse gapless modes dispersion at least as soft as $|\bm{k}|^2$ at small $\bm{k}$.

\section{The effective field theories}
\label{sec: field theory}

\subsection{The RK point}

A field theory can well describe the long-wavelength physics at the RK point~\cite{PhysRevB.69.224415,PhysRevB.69.224416,moessner2010quantum,fradkin2013field}. To define the theory, we introduce a vector field $\bm{d}(\bm{r})$ which is the coarse-grained $\tau_{ij}/2$ in the sense that for any large enough surface $S$,
\begin{align}
    \sum_{\tau_{ij}\in S} \tau_{ij}/2 = \int d \bm{S} \cdot \bm{d}
\end{align}
In this convention, each defect thus generates a unit flux of $\bm{d}$, and each component of $\bm{d}$ has a maximal value $d_\text{max}\equiv 1/a_0^2$ where $a_0$ is the lattice constant. 

The action of the field theory in Euclidean spacetime reads:
\begin{align}\label{eq: RK field theory}
    S[\bm{d},\bm{a}] = \int d\tau d^3\bm{r} \left[\mathrm{i}\bm{d} \cdot \dot{\bm{a}} + \frac{1}{2\tilde{\mu}}\left(\kappa^2|\nabla\times\bm{d}|^2 +|\nabla \times \bm{a}|^2 \right)\right] 
\end{align}
where $\Tilde{\mu}$ is a constant, and $\bm{a}$ is the canonical conjugate variable of $\bm{d}$ in the sense that they obey the canonical commutation relation on a time slice after quantization (we put hats on the variables to represent their operators):
\begin{align}
    [\hat{\bm{d}}_i(\bm{r}), \hat{\bm{a}}_j(\bm{r}')] = \mathrm{i} \delta(\bm{r}-\bm{r}') \delta_{ij}
\end{align}
$\kappa$ is the constant that appeared in the correlation function Eq.~\ref{eq: classical correlation function}. To see that this indeed captures all the essential physics of the model at long wavelength, we first diagonalize the action in the momentum and imaginary time coordinate
\begin{align}
     S[\bm{d},\bm{a}] = \int d\tau  \frac{d^3 \bm{k}}{(2\pi)^3} \left[\mathrm{i}\bm{d}(\bm{k}) \cdot  \dot{\bm{a}}(-\bm{k}) + \frac{ |\bm{k}|^2 }{2\tilde{\mu}}  \left(  \kappa^2 |\bm{d}_\perp (\bm{k})|^2  +  | \bm{a}_\perp(\bm{k})|^2\right) \right]
\end{align}
where $\bm{d}_\perp$ and $\bm{a}_\perp$ are the components of $\bm{d}$ and $\bm{a}$ perpendicular to $\bm{k}$. The independence of the action on the parallel component of $\bm{a}$, $\bm{a}_\parallel$, suggests that it should be regarded as a Lagrangian multiplier at all imaginary time $\tau$. Integrating out this multiplier, we find that this enforces a constraint on the Hilbert space of $\bm{d}$:
\begin{align}
   \bm{d}_\parallel \propto \bm{d}(\bm{k}) \cdot \bm{k} = 0 \implies  \nabla \cdot \bm{d}(\bm{r}) = 0.
\end{align}
This is exactly the coarse-grained version of the ice rule constraint.

Further integrating out $\bm{a}_\perp$, we obtain an action for $\bm{d}$ field that is diagonal in momentum-Matsubara frequency coordinate:
\begin{align}
     S[\bm{d}] = \frac{1}{2} \sum_{\Omega_n = 2\pi n T } \int \frac{d^3 \bm{k}}{(2\pi)^3} \left(  \kappa^2 |\bm{k}|^2 /\tilde{\mu} +  \tilde{\mu} \Omega_n^2/|\bm{k}|^2 \right) |\bm{d}_\perp (\bm{k},\Omega_n)|^2 
\end{align}
The propagator of the two transverse modes has poles with a dispersion relation
\begin{align}
    E(\bm{k}) = \frac{\kappa}{\tilde{\mu}} |\bm{k}|^2
\end{align}
which agrees with the prediction of the SMA analysis.

Lastly, integrating out the $\Omega_n$ dependence in the inverse kernel at zero temperature ($T\rightarrow 0$), we reproduce the structure factor of the equal-time correlator in Eq.~\ref{eq: classical structure factor}:
\begin{align}
   S_{\alpha\beta}(\bm{k}) = \langle \bm{d}_{\alpha} (-\bm{k}) \bm{d}_{\beta}(\bm{k})\rangle/(2\pi)^3 = 2 \lim_{T\rightarrow 0} T \tilde{\mu} \sum_{\Omega_n} \frac{|\bm{k}|^2-\bm{k}_\alpha\bm{k}_\beta}{\kappa^2  |\bm{k}|^4 +  \tilde{\mu}^2 \Omega_n^2} = \frac{1}{\kappa} \left(1-\bm{k}_\alpha \bm{k}_\beta/|\bm{k}|^2\right).
\end{align}
Note that $\kappa = 4\kappa_\tau$ due to our convention $\tau /2 \sim \bm{d}$.

Therefore, we conclude that this is a correct field theory for the long-wavelength physics of the system. We mention that this action can alternatively be `derived' from a field theory of the classical ice-rule vertex model at infinite temperature, following the recipe in Ref.~\cite{moessner2010quantum}. 

\subsection{The vicinity of the RK point}

In the vicinity of the RK point, one should take other possibly relevant terms into consideration in the effective field theory. The result is:
\begin{align}\label{eq: RK vicinity field theory}
     S[\bm{d},\bm{a}] = \int d\tau d^3\bm{r} \left[\mathrm{i}\bm{d} \cdot \dot{\bm{a}} + \frac{1}{2\tilde{\mu}}\left(\kappa^2|\nabla\times\bm{d}|^2 +|\nabla \times \bm{a}|^2 \right)+ \frac{1}{2\tilde{\epsilon}}|\bm{d}|^2 + \frac{c_4}{4} |\bm{d}|^4  + \frac{c_4'}{4} (\bm{d}_x^4+\bm{d}_y^4+\bm{d}_z^4) + \dots \right] 
\end{align}
It turns out that the RK point is always a critical point since the term $|\bm{d}|^2$ is always relevant; the last two terms in the above expressions are dangerously irrelevant at the RK point since they will become relevant as soon as $1/\Tilde{\epsilon}$ becomes negative~\cite{PhysRevB.69.224415}. Moreover, in contrast to the two-dimensional cases, there the integer nature of $\bm{d}$ is irrelevant~\cite{PhysRevB.69.224415,PhysRevB.69.224416} so that no incommensurate crystallization would be stabilized.  This transition point separates two phases corresponding to the two signs of $\Tilde{\epsilon}$, as we will discuss below.

\subsubsection{$\Tilde{\epsilon}>0$: $U(1)$ gauge field}

When $\Tilde{\epsilon}>0$, the field theory at long-wavelength becomes (dropping irrelevant terms)
\begin{align}\label{eq: U(1) gauge field theory}
     S_{U(1)}[\bm{d},\bm{a}] = \int d\tau d^3\bm{r} \left[\mathrm{i}\bm{d} \cdot \dot{\bm{a}} + \frac{1}{2\tilde{\mu}}|\nabla \times \bm{a}|^2 + \frac{1}{2\tilde{\epsilon}}|\bm{d}|^2 \right] 
\end{align}
We immediately find that this is exactly the action of a $3+1$D $U(1)$ gauge theory in the Weyl (temporal) gauge as we identify $\bm{d}$ to be the electric displacement field ($\bm{d}=\tilde{\epsilon} \bm{e}$ is proportional to the electric field), $\bm{a}$ to be the vector potential (which determines the magnetic flux density $\bm{b}=\nabla \times \bm{a}$, and thus the magnetic field $\bm{h} = \bm{b}/\Tilde{\mu}$). Note that, we have used lowercase symbols to distinguish them from the native electromagnetic (EM) fields. The existence of this phase in the vicinity of the RK point has been numerically verified for a similar model on diamond lattice~\cite{PhysRevLett.108.067204,PhysRevLett.127.117205,PhysRevB.84.115129}.

The properties of the theory are well known. It features two linear gapless `photon' excitations with dispersion $E^{\text{photon}}(\bm{k}) =\tilde{c} $ where the `speed of light' $\tilde{c} =1/\sqrt{\tilde{\epsilon}\tilde{\mu}}$. Since each defect in the system carries one unit flux of $\bm{d}$, it carries charge $1$ under the emergent gauge field. Since no fractional defects are definable in this system, this emergent $U(1)$ gauge field is compact with elementary charge $\tilde{e}=1$. Two elementary emergent charges separated by $\bm{r}$ interact with potential
\begin{align}
    V(\bm{r}) = \frac{1}{4\pi \tilde{\epsilon} |\bm{r}|}
\end{align}
Since this is not the EM field of the universe, magnetic monopole excitations are also possible, which have elementary magnetic charge $\tilde{g} = 2\pi/\tilde{e}$ due to Dirac quantization. They also interact with a $1/r$ potential.

Then, the fine structure constant of the emergent gauge field is (setting $\hbar =1$)
\begin{align}
    \tilde{\alpha} = \frac{1}{4\pi \tilde{\epsilon}\tilde{c}} = \frac{1}{4\pi} \sqrt{\frac{\tilde{\mu}}{\tilde{\epsilon}}}
\end{align}
When this constant exceeds $\sim 0.2$, it is conjectured that this theory confines on lattices~\cite{CARDY1980369, LUCK1982111, PhysRevD.56.3896}. This could only happen when the system is away from the RK point since $1/\tilde{\epsilon}$ vanishes exactly at the RK point.

In this `quantum Coulomb' phase, at zero temperature, the structure factor of the equal-time correlator becomes (in place of the `classical Coulomb' result in Eq.~\ref{eq: classical structure factor}):
\begin{align}
   S_{\alpha\beta}(\bm{k}) = \langle \bm{d}_{\alpha} (-\bm{k}) \bm{d}_{\beta}(\bm{k})\rangle/(2\pi)^3 = \frac{|\bm{k}|}{4\pi \tilde{\alpha} } \left(1-\bm{k}_\alpha \bm{k}_\beta/|\bm{k}|^2\right).
\end{align}
which leads to a $\sim 1/|\bm{r}|^4$ decaying correlation function.

So far, we have neglected the background EM gauge field. However, since both the emergent gauge field and the native EM gauge field are in the same dimension, they should both be viewed as dynamic. We thus consider the full theory containing both the emergent $U(1)$ gauge field and the native EM field in our universe (represented by uppercase letters):
\begin{align}\label{eq: U(1) gauge field theory + background}
     S[\bm{d},\bm{a};\bm{E},\bm{A}] = \int d\tau d^3\bm{r} &\left[ \mathrm{i}\bm{d} \cdot \dot{\bm{a}} + \frac{1}{2\tilde{\epsilon}}|\bm{d}|^2 + \frac{1}{2\tilde{\mu}}|\nabla \times \bm{a}|^2  \right.\nonumber\\
     &\ \ \ + \mathrm{i}\epsilon \bm{E} \cdot \dot{\bm{A}}  + \frac{\epsilon}{2}|\bm{E}|^2 + \frac{1}{2\mu}|\nabla \times \bm{A}|^2 \nonumber\\
     &\ \ \ \ \ \ - \eta \bm{d} \cdot \bm{E} +\frac{\eta}{\tilde{\mu}} (\nabla \times \bm{a})\cdot (\nabla \times \bm{A})\Big{]} 
\end{align}
where $\eta$ is the magnitudes of electric dipole moment carried by each displaced bond normalized by the lattice constant. 

The couplings between native and emergent gauge fields consist of two parts:
\begin{align}
-\bm{P}\cdot \bm{E} - \bm{M} \cdot \bm{B}
\end{align}
where the polarization $\bm{P}$ is nothing but $\eta \bm{d}$ and the magnetization $\bm{M}$ is determined by $\nabla\times \bm{M} = -\mathrm{i}\eta \dot{\bm{d}}$ since the motions of the microscopic dipoles are the only source of the magnetization ($-\mathrm{i}$ factor comes from Wick rotation since we are working in Euclidean space). After considering the equation of motion of $\bm{a}$: $\mathrm{i}\dot{{\bm{d}}} = \nabla\times\left(\nabla\times \bm{a} \right)/\tilde{\mu}$, we see that $\bm{M} = -\eta \nabla\times\bm{a} /\tilde{\mu}$. So the second term further equates $+\frac{\eta}{\tilde{\mu}} (\nabla \times \bm{a})\cdot (\nabla \times \bm{A})$.

Various properties of the combined system can be solved. For example, by solving the poles of the propagator, we find that the emergent photon and the native photon mix into two branches with renormalized speeds of light
\begin{align}
    c_{+}^2  = c^2-\frac{\eta^2}{\epsilon\tilde{\mu}} \ , \ \ c_-^2 = \tilde{c}^2-\frac{\eta^2}{\epsilon\tilde{\mu}}
\end{align}
We note that when $\frac{1}{\tilde{\epsilon}\eta^2}<\frac{1}{\epsilon}$, $c_-^2$ becomes negative, which implies that the emergent $U(1)$ gauge field becomes unstable. This can also be seen by the divergence of the static, uniform susceptibility of $\bm{E}$. We note that since the RK point corresponds to $\frac{1}{\tilde{\epsilon}}=0$, this means that once the coupling to EM gauge fields is considered, the RK point is no longer a transition point of the RVB phase. 

Moreover, the coupling between the emergent and native gauge fields suggests that the charge and monopole excitations of the emergent gauge fields are also charged under the EM field. Specifically, one emergent charge excitation emits one unit flux of $\bm{d}$, which induces $\eta$ unit flux of $\bm{E}$; therefore, each charge excitation carries charge $\eta$ under the EM field. Similarly, each emergent monopole emits one unit flux of $\bm{b}=\nabla\times \bm{a}$, which induces $\eta /\tilde{\mu}$ unit flux of $-\bm{M}$ and $\bm{H}$ and therefore carries magnetic charge $\eta /\tilde{\mu}$.

\subsubsection{$\Tilde{\epsilon}<0$: Ferroelectricity$^*$}

When $\Tilde{\epsilon}<0$, It is necessary to include the higher order terms in $|\bm{d}|$ in the potential energy of $\bm{d}$:
\begin{align}
     S[\bm{d},\bm{a}] =& \int d\tau d^3\bm{r} \left[\mathrm{i}\bm{d} \cdot \dot{\bm{a}} + \frac{1}{2\tilde{\mu}}\left(\kappa^2|\nabla\times\bm{d}|^2 +|\nabla \times \bm{a}|^2 \right) + \mathcal{E}(\bm{d}) \right]  \\
     \mathcal{E}(\bm{d}) =&   -\frac{1}{2|\tilde{\epsilon}|} |\bm{d}|^2 + \frac{c_4}{4} |\bm{d}|^4  + \frac{c_4'}{4} (\bm{d}_x^4+\bm{d}_y^4+\bm{d}_z^4) 
\end{align}
The saddle-point solutions to this action have static and uniform $\bm{b}_\text{saddle}$. The solution reads (which can be readily related to the other symmetry-equivalent ones):
\begin{align}
    \bm{b}^\star = \begin{cases}
        \frac{1}{\sqrt{|\tilde{\epsilon}| (c_4+c_4')}} \left(1,0,0\right) & \text{ if } c_4'< 0 \\
        \frac{1}{\sqrt{|\tilde{\epsilon}| (3c_4+c_4')}} \left(1,1,1\right) & \text{ if } c_4'> 0 
    \end{cases}
\end{align}

When $1/|\Tilde{\epsilon}|$ is small  (i.e. not too far from the RK point) in the sense that each component of $\bm{b}_\text{saddle}$ is small compared with $d_\text{max} =1/a_0^2 $, the expansion of potential in powers of $\bm{d}$ is justifiable, and the saddle-point solution is a partial-ferroelectric state. The fluctuation around these solutions can be captured by expanding the action around the saddle -point, which leads to (defining $\delta\bm{d} \equiv \bm{d}-\bm{d}_\text{saddle}$):
\begin{align}
    S^\star[\delta\bm{d} ,\bm{a}] =& \int d\tau d^3\bm{r} \left[\mathrm{i}\delta \bm{d} \cdot \dot{\bm{a}} + \frac{1}{2\tilde{\mu}}|\nabla \times \bm{a}|^2  + 
\sum_{\alpha,\beta = x,y,z}\frac{1}{2}(\tilde{\epsilon}^\star)^{-1}_{\alpha\beta} \delta \bm{d}_\alpha \delta \bm{d}_\beta \right]
\end{align}
where 
\begin{align}
    (\tilde{\epsilon}^\star)^{-1}_{\alpha\beta} \equiv \left.
 \frac{\partial^2 \mathcal{E}(\bm{d}) }{\partial \bm{d}_\alpha \partial \bm{d}_\beta}\right|_{\bm{d} = \bm{d}_\text{saddle}}
\end{align}
is an effective dielectric constant tensor. This action still describes a dynamical $U(1)$ (albeit anisotropic) gauge field. Therefore, the system corresponding to this parameter regime has coexisting partial ferroelectric order and gauge fields, which we dub the ``Ferroelectricity$^*$'' (FE$^*$) phase. Note that, when $1/|\tilde{\epsilon}|$ gets large, this phase will eventually be fully FE ordered. The monopole confinement might further facilitate this as the effective fine structure constant $\sim{\tilde{\mu}/|\tilde{\epsilon}|}$ gets large.


\section{A possible phase diagram}

\begin{figure}[t]
    \centering
    \includegraphics[width = 0.4\linewidth]{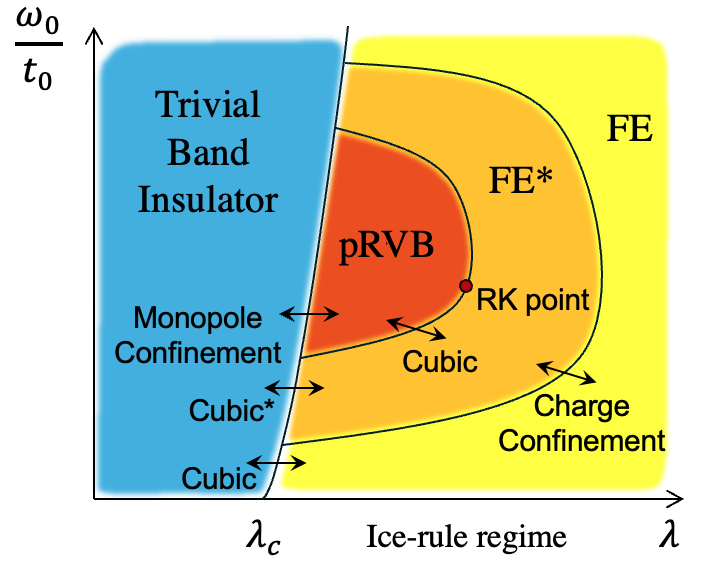}
    \caption{A possible phase diagram of the model in Eq.~\ref{eq: microscopic model}.}
    \label{fig: phase diagram}
\end{figure}

Putting the analyses in the previous sections together, we now propose possible phase diagrams of the model in Eq.~\ref{eq: microscopic model}. The reasoning is as follows. For sufficiently small $\lambda =\Ueph/t_0$, the system is a trivial insulator, whereas when it exceeds a certain value, the ice rule emerges as a constraint on the low energy subspace. The boundary between these two regimes is set by the condition $S\sim (\lambda-\lambda_c)^{3/2} t_0/\omega_0 \gg 1$ where $S$ is the action of the instanton tunneling event between two ice-rule configurations that differ by a plaquette flip (See Sec.~\ref{sec: scaling}), therefore the boundary shifts towards larger $\lambda$ side as $\omega_0$ increases. Within the ice-rule regime, various ordered phases and the pRVB state are possible. Assuming that the additional terms will eventually drive the system to a classical ferroelectric phase, a particularly interesting possibility is that the transition can go through intermediate phases pRVB and FE$^*$.

The nature of the transitions between different phases can be deduced based on the symmetry-breaking pattern and the fate of excitations. The simplest possibility is the transition between the trivial band insulator and FE phases, which is in the cubic universality class for cubic lattice. However, the transition between the trivial band insulator and pRVB phases does not involve any ($0$-form) symmetry, whereas the charge excitations of the emergent gauge field (the defects violating the ice rule) condense at the transition since they have vanishing creating energy there, leading to a Higgs (or equivalently monopole confinement) transition~\cite{PhysRevD.19.3682,thorngren2023higgs}. Similarly, the transition between FE and FE$^*$ may be associated with no ($0$-form) symmetry breaking but is driven by a charge confinement transition due to monopole condensation (due to the large fine structure constant). At the transition from FE$^*$ to the trivial band insulator, the recovery of cubic symmetry could be understood as driven by the condensation of charge excitations; we thus call this simultaneous symmetry breaking and gauge transition a cubic$^*$ transition, following the nomenclature in existing literature~\cite{PhysRevB.109.L241109}.

\section{Phenomenological signatures of the pRVB phase}

Multiple novel phenomena are associated with the pRVB phase and the ice rule,  which we will discuss in this section. Some of the aspects have been considered for water ice~\cite{PhysRevB.93.125143,PhysRevB.74.024302,PhysRevB.99.174111}

\subsection{Consequences of the ice rule}

\subsubsection{Phonon spectrum in the trivial phase}

As we have seen in Sec.~\ref{sec: implication on phonon spectrum}\&\ref{sec: cubic numerics}, there are two exactly flat transverse optical phonon modes in the trivial band insulator phase, whose flatness is tied to the special hopping matrix and thus the exactness of the ice rule at the larger coupling. This feature can be used as a precursor for the emergence of the ice rule before a material really enters such a regime.

\subsubsection{Residual entropy}

When the ice rule does emerge, it leads to large residual entropy detectable at a temperature low compared to the normalized optical phonon frequency but high compared to all the other emergent energy scales. The value of the entropy density can be estimated with Pauling's argument in Sec.~\ref{sec: exact ice rule}.

\subsubsection{Pinch points in the correlation functions}

The ice rule (or other types of constraints on the Hilbert space) in general can be written as~\cite{PhysRevB.71.014424,PhysRevLett.127.107202,PhysRevB.88.024407,PhysRevB.91.245152} 
\begin{align}
    \bm{L}^{(i)}(\bm{k}) \cdot \bm{u}_{\bm{k}} = 0 
\end{align}
in momentum space, where $\bm{L}^{(i)}(\bm{k})$ are the constraint vectors, and  $\bm{u}_{\bm{k}}$ is the Fourier transform of all degrees of freedom within each unit cell. In the current example of cubic lattice ice rule, $\bm{L}(\bm{k}) = \bm{k}$ and  $\bm{u} = (u_x, u_y, u_z)$. This constraint restricts the form of the correlation functions $\langle \bm{u}_\alpha \bm{u}_\beta\rangle$ to have a singularity whenever some of $\bm{L}^{(i)}(\bm{k})$ vanishes. Those points are called pinch points, which should be visible in the scattering measurement of correlation functions. 

\subsection{Signatures of the pRVB phase}

\subsubsection{Phonon spectrum}

As seen in Sec.~\ref{sec: field theory}, one of the Hallmarks of the pRVB phase is the existence of emergent photon modes. Those modes are eventually the highly renormalized TO phonon modes, which are measurable through neutron scattering experiments. 

The propagation velocity of these phonon modes, $\tilde{c}$, will become arbitrarily slow at the boundary between pRVB and FE$^*$ phase since it corresponds to the RK point where $1/\tilde{\epsilon}$ vanishes.

\subsubsection{Deconfined excitation with irrationally fractionalized electric charge}

As seen in Sec.~\ref{sec: field theory}, the charge excitations of the emergent gauge field are deconfined and interact with a Coulomb potential. Each excitation corresponds to a local violation of the ice rule, and thus carries charge $|Q| = \eta/a_0$ under the EM field where $\eta$ is the dipole moment amplitude of each displaced bond and $a_0$ is the lattice constant. As seen in Sec.~\ref{sec: scaling}, $\eta$ depends on all sorts of microscopic details on the parameters of the system, thus it generically is a small and irrational number. We note that similar notion has been put forward for the water ice~\cite{PhysRevB.91.245152}.

Two charge excitations of opposite charges attract with an attractive Coulomb potential, thus such a pair can form a bound state. This will be a novel form of `exciton' excitation of a band insulator.

\subsubsection{Conductivity}

The pRVB phase has a large electronic gap $\epsilon_\text{gap} \sim \mathcal{O}(t_0)$, as seen in Sec.~\ref{sec: cubic numerics}. One would naively expect no conductivity below this gap. However, the charge excitations discussed above carry an electric charge and are deconfined so that they can conduct electric current whenever they are created. Therefore, one should see finite conductivity when either the frequency or the temperature is greater than the creation energy of such excitaions, $E^\text{defect}$. As discussed in Sec.~\ref{sec: scaling}, this energy scale depends on the distance to the trivial band insulator phase, which can be arbitrarily low. Therefore, the pRVB phase has much larger conductivity than any other surrounding phases despite the persistence of a large electronic gap.  This is a spectacular form of “superionic conduction.”

\section{Versions of the model suitable for Determinantal Quantum Monte Carlo}

The model we have introduced - in common with most models of the electron-phonon problem - is amenable to solution by sign-free determinantal quantum Monte Carlo (DQMC).  This means that, at a future date, it should be possible to explore the phase diagram of this model away from the particular limits on which we have focussed to obtain analytical control.  Thus, the robustness of the RVB phase - among other things - can be addressed in this manner.

The model is still challenging from a numerical perspective because one needs to sample the continuous space-time configurations of the phonon fields.  Thus, for the purposes of DQMC, it may be easier to replace the model with a simplified version in which the phonon mode $X_{ij}$ is represented by a two-state pseudo-spin $\tau_{ij}$.  Here $\tau_{ij}^z=-\tau_{ji}^z=\pm 1$ is a proxy for $X_{ij} = \pm x_0$, and a transverse field (coupled to $\tau_{ij}^x$) is introduced to represent the phonon dynamics.  The resulting simplified model is
\begin{align}\label{eq:DQMCmodel}
    \hat{H} = \sum_{i}\Delta \hat{n}_i - t\sum_{\langle ij\rangle, \sigma} \left(\hat{c}^{\dagger}_{i\sigma} \hat{f}_{\langle ij \rangle  \sigma}
    + \hat{c}^{\dagger}_{i\sigma} \hat{f}_{\langle ij \rangle  \sigma}+\text{h.c.}\right)
    - \alpha\sum_{\langle ij\rangle, \sigma} \tau_{ij}^z\left(\hat{c}^{\dagger}_{i\sigma} \hat{f}_{\langle ij \rangle  \sigma}
    -\hat{c}^{\dagger}_{i\sigma} \hat{f}_{\langle ij \rangle  \sigma}+\text{h.c.}\right)-h \sum_{\langle ij \rangle}  \tau_{ij}^x
\end{align}
Another advantage of this model is that it has relatively few dimensionless parameters ($\Delta/t$, $\alpha/t$, and $h/t$).  It is in the trivial phase for large $h/t$, and in an appropriate classical broken symmetry phase for small $h/t$.  The interesting phases - and presumably the RVB phase - should arise for intermediate $h/t$ in various regions of the $\Delta -\alpha$ plane.

\bibliography{ref}